\documentclass[natbib]{svjour3}                     
\smartqed  
\usepackage{graphicx}
\usepackage{caption}
\usepackage{subcaption}
\usepackage{amsmath}
\usepackage{textcomp}

\newcommand{\aap}{{Astron. Astrophys.}}

\begin{document}
\title{ExoSim: the Exoplanet Observation Simulator
}

\titlerunning{ExoSim: the Exoplanet Observation Simulator}        

\author{Subhajit Sarkar {$\cdot$} Enzo Pascale {$\cdot$} \\ Andreas Papageorgiou {$\cdot$}  Luke J. Johnson {$\cdot$} \\ Ingo Waldmann}

\authorrunning{S.Sarkar et. al} 

\institute{S. Sarkar, A. Papageorgious \at
              School of Physics and Astronomy, Cardiff University, The Parade, Cardiff CF24 3AA, U.K.\\
              \email{subhajit.sarkar@astro.cf.ac.uk}           
           \and
           E. Pascale \at
              La Sapienza University of Rome, Department of Physics, Piazzale Aldo Moro 2, 00185, Rome, Italy.\\           
          \and 
           L.J. Johnson \at
              Department of Physics, Imperial College London, London, SW7 2AZ, U.K.\\                  
           \and       
           I. Waldmann \at
              Department of Physics and Astronomy, University College London, Gower St, London WC1E 6BT, U.K.
}

\date{Received: date / Accepted: date}

\maketitle

\begin{abstract}

A new generation of exoplanet research beckons and with it the need for simulation tools that accurately predict signal and noise in transit spectroscopy observations.  We developed ExoSim: an end-to-end simulator that models noise and systematics in a dynamical simulation. ExoSim improves on previous simulators in the complexity of its simulation, versatility of use and its ability to be generically applied to different instruments. It performs a dynamical simulation that can capture temporal effects, such as correlated noise and systematics on the light curve. It has also been extensively validated, including against real results from the Hubble WFC3 instrument.  We find 
ExoSim is accurate to within 5\% in most comparisons. ExoSim can interact with other models which simulate specific time-dependent processes. A dedicated star spot simulator allows ExoSim to produce simulated observation that include spot and facula contamination. ExoSim has been used extensively in the Phase A and B design studies of the ARIEL mission, and has many potential applications in the field of transit spectroscopy.

\end{abstract}

\section{Introduction}
\label{intro}
Today thousands of exoplanets have been confirmed, revealing a diverse population in size, mass, temperature and orbital properties.  The characterization of exoplanets through spectroscopic analysis of their atmospheres is key to fully understanding the properties of the individual planets. Such observations can help to constrain theories of planet migration, formation and evolution, shedding light on the mechanisms underlying the great diversity seen in the exoplanet population. 

Transit spectroscopy, theorized by \cite{Seager2000} and first demonstrated by \cite{Charbonneau}, has been the major technique used to obtain exoplanet spectra to date.  This technique returns the transmission spectrum at the day-night terminator. It has been used to discover atmospheric atomic species such as sodium \citep{Charbonneau, Redfield2008}, potassium \citep{Colon2012}, hydrogen \citep{Vidal2003} and helium \citep{Spake2018}, hazes \citep{Pont2008} and clouds \citep{Kreidberg2014, Knutson2014}, and molecular species including water \citep{Tinetti2007, Sing2016}.  The similar technique of eclipse spectroscopy returns the dayside emission spectra of the exoplanet, and can provide constraints on the vertical temperature-pressure profile of the atmosphere. Eclipse spectra have detected molecular species including water, methane, CO and CO$_2$ \citep{Grillmair2008, Swain2009a, Swain2009b}.  In phase-resolved emission spectroscopy \citep{Stevenson2014,Arcangeli2019}, multiple emission spectra are obtained as a function of phase.

Both transit and eclipse spectroscopy  operate in the time domain, relying on  high precision spectrophotometric light curve measurements.  These require high levels of photometric stability to be maintained over the time scale of the planet transit or eclipse. The light curves are used to extract fractional transit depths at different wavelengths.  The wavelength-dependent variations in these transit depths trace out the exoplanet spectrum. In primary transit, the transit depth gives $(R_p/R_s)^2$, where $R_p$ is the apparent radius of the planet, and $R_s$ is the radius of the star. In secondary eclipse, the eclipse depth gives $F_p/F_s$, i.e. the contrast ratio, where $F_p$ is the flux from the planet, and $F_s$ is the flux from the star. 

These wavelength-dependent variations in transit or eclipse depths can give spectral amplitudes in the order of a few tens to hundreds of ppm of the stellar flux, depending on the planet in question. The detection of such small signals is thus highly vulnerable to noise and systematics.  Determination of the experimental uncertainties require accounting for at a similar level of precision.  This is complicated by the time domain nature of the observation which make it vulnerable to the effects of time-correlated (`coloured') noise and time-dependent systematics that may distort the light curve. Estimating the correct experimental uncertainties on the final light curve measurements and the emergent spectrum is essential for confidence of the final scientific conclusions. 

To address the need for better estimation of experimental precision and accuracy, ExoSim, the Exoplanet Observation Simulator, was developed as a generic end-to-end simulator of transit spectroscopy observations.  ExoSim is publicly available on GitHub\footnote{https://github.com/ExoSim/ExoSimPublic}.ExoSim is designed to be flexible, versatile and applicable to different instruments.  ExoSim operates dynamically, modelling the time domain directly in small steps. This gives it the potential to better capture the effects of correlated noise and systematics on the final spectrum.  Dynamical simulation permits the investigation of complex noise sources such as pointing jitter or stellar activity on an observation.  This gives ExoSim the potential to investigate the performance of both new and established instruments under complex conditions, and thus optimize the scientific potential of space missions.

\section{ExoSim}

ExoSim draws on the experience gained from the development and application of EChOSim \citep{Pascale2015}, a simulator developed for the EChO mission concept \citep{Tinetti2012}.  It adopts a similar modular structure and nomenclature. ExoSim however improves upon EChOSim in several ways, both in function and structure.  The algorithmic differences are described in this paper. Although EChOSim was capable of simulating different instruments, the ease of changing the instrumental parameters is improved in ExoSim.  ExoSim has also been more comprehensively validated than EChOSim.

The development of ExoSim has been driven by the Atmospheric Remote-sensing Exoplanet Large-survey (ARIEL) mission \citep{Tinetti2018}, with the goal of producing a reliable, realistic and accurate end-to-end simulator. ARIEL is a 0.9 m space telescope which will perform the first space-based and large scale survey of about 1000 exoplanets.  The simulator was needed to direct and test design iterations of the science payload, and perform simulations to assess the impact of complex noise sources  on scientific return.  ExoSim was used extensively in the performance evaluation of ARIEL during its Phase A study \citep{Sarkar2017}.  However ExoSim has been designed to be generic. This feature  not only extends its use to other observatories, but also allows it to be validated against existing instruments which are already producing data. 

In this paper we describe the ExoSim algorithm, highlighting where it differs from that previously described for EChOSim.
We also present validation testing results, firstly comparing ExoSim signal and noise against equations. The pointing jitter model is validated against an analytic expression and a simple independent simulation. We then perform a cross-validation of ExoSim with an independently created radiometric simulator, the ESA Radiometric Model \citep{Tinetti2018}.  Finally we compare ExoSim focal plane counts and simulated light curves with those from published studies using the Hubble Wide Field Camera 3 (WFC3) IR instrument.

\section{Overview}
\label{sec:1}

ExoSim models the host star and planet transit event, simulating  the temporal change in stellar flux due to this, i.e. the light curve.  It does this in a wavelength-dependent manner, using an input planet spectrum to determine the light curve depth for any given wavelength. It also models the optical system, consisting of the telescope (`common optics'), one or more instrument channels, and their associated detectors.  The instrument channels can be spectroscopic or photometric.  ExoSim simulates the modification of the signal as it passes through the optical system, and generates a focal plane image of the star, either as a photometric or a spectral image.  ExoSim simulates an observation as a series of exposures consisting of up-the-ramp non-destructive reads (NDRs).  The detectors are assumed to be mercury-cadmium-telluride infra-red detectors \citep[e.g.][]{Jerram2019}. Noise and systematics are simulated and added to the images. This image time series thus contains the planet transit event, as well as the effects of numerous noise sources and systematics.  
The inputs to ExoSim are an Input Configuration File, which contains user-defined parameters for the observation and instrument. This in turn references several instrument specific reference files. The final output is a data cube of NDRs packaged in FITS format. This requires processing by a data reduction pipeline. The simulations can be run without a transit event (`out-of-transit'), with various noise sources switched on or off, and can also be run as a Monte Carlo simulation to obtain a probability distribution of transit depths as a way of finding the `error bar' on the spectrum.  The data pipeline is not part of ExoSim, but is required to process the ExoSim output, extract the required signal and noise information, and reconstruct the planet spectrum.

\section{Algorithm}

\begin{figure}
\begin{center}
\includegraphics[trim={0, 7cm, 0, 0}, clip,width=0.9\textwidth]{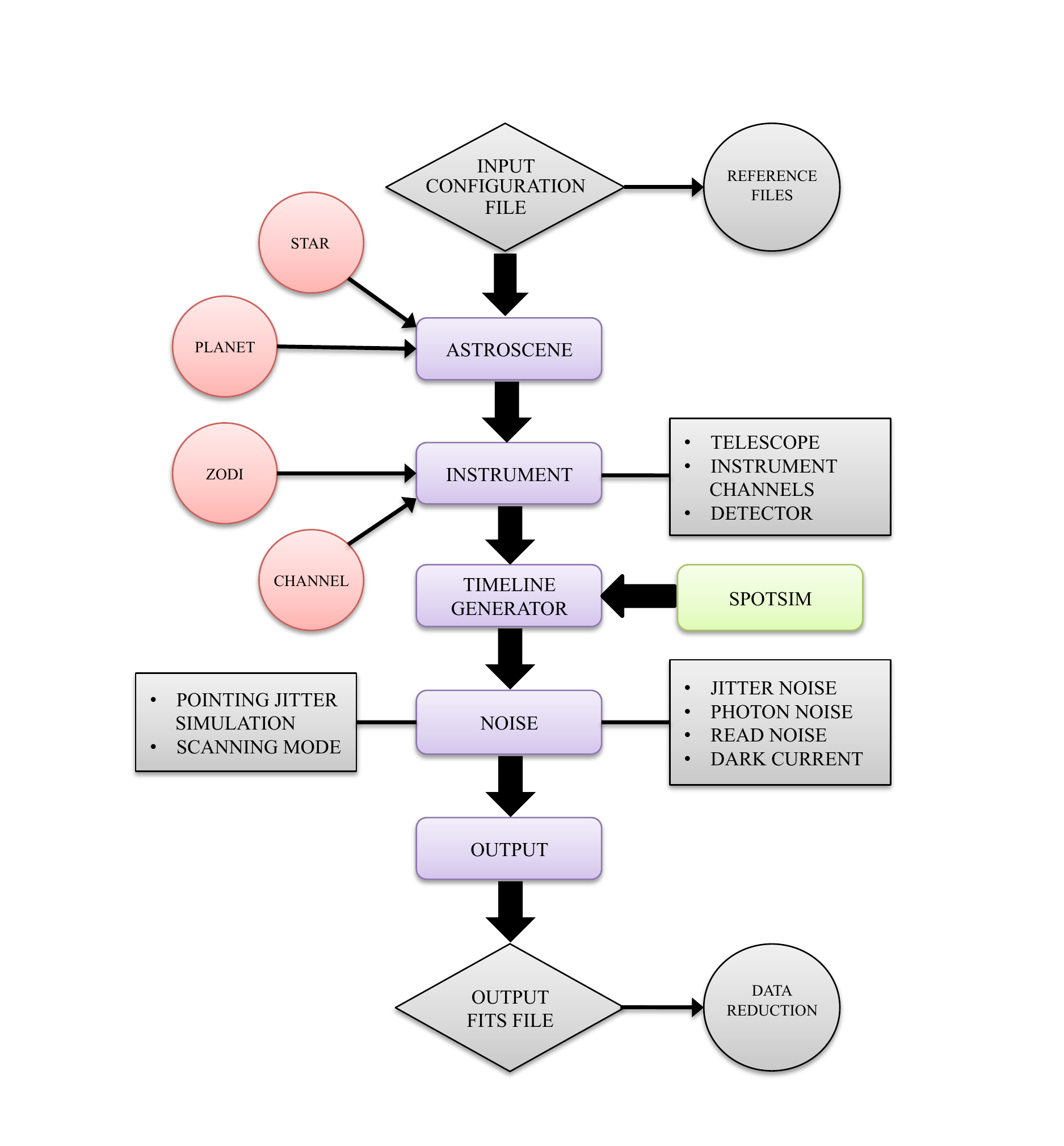}
\caption[]{ExoSim modular architecture and informational flow. The core of the algorithm is completely generic. The Input Configuration File sets the simulation parameters and calls on instrument-specific reference files. Modules shown in purple can be upgraded. Object classes are shown in red circles. SpotSim, a dedicated star spot simulator, interacts with ExoSim in the \textit{Timeline generator} module.}
\label{Chapter2:Figure:architecture}
\end{center}
\end{figure}

The ExoSim algorithm simulates a physical model which is described within the modular structure of the code (Figure \ref{Chapter2:Figure:architecture}) as follows.

\subsection{Astroscene}

The host star radiates isotropically with a surface flux density, $F_s(\lambda)$.  After reaching the telescope aperture at a distance $D$, the flux density falls to $F_{tel}(\lambda)=F_s(\lambda)(R_s/D)^2$. The \textit{Astroscene} module instantiates \textit{Star} and \textit{Planet} object classes.  The \textit{Star} class contains a PHOENIX stellar model \citep{Allard} matched to the host star parameters, and obtains $F_{tel}(\lambda)$. 

As the planet transits the star, the flux from the star is modulated in time forming a light curve.  The shape of the light curve depends on: $z$ (the time grid of the projected distance between the centre of the star and the centre of the planet in units of the star radius), stellar limb-darkening, and the planet-star radius ratio. The latter two are wavelength-dependent parameters.  In eclipse, the light curve depends on $z$ and the contrast ratio between the star and the planet. $z$ is a function of the period, the semi-major axis, the star radius, orbital inclination, eccentricity, and time.  The \textit{Planet} class calculate $z$ from the orbital parameters, and generates wavelength-dependent light curves using the formulation of \cite{Mandel}. An input planet spectrum is used to give the planet-star radius or flux ratios. This can be produced from third party radiative transfer models such as TauREx \citep{Waldmann}, NEMESIS \citep{Irwin} or CHIMERA \citep{Line2013}.

Compared to EChOSim, ExoSim can now select star and planet parameters automatically from databases given a chosen planet identified in the Input Configuration File.  ExoSim can also generate `integrated' light curves, where the variation of the light curve within the integration time of an image is accounted for. This may be more accurate for long integrations than using `instantaneous' light curves, where the value of the light curve at an instant of time is applied. Compared to EChOSim, ExoSim uses improved limb-darkening coefficients generated from ATLAS and PHOENIX models \citep{Morello2017}.

An additional astrophysical source of photons is the zodiacal light, which is modeled in the \textit{Zodi} object class. The zodiacal light is a diffuse source (rather than a point source). ExoSim utilizes a zodi model based on \cite{Glasse2010} and 
is processed in a similar way to that described in \cite{Pascale2015}.

\subsection{Instrument}

\subsubsection{Telescope}

The starlight then enters the telescope aperture of area $A_{tel}$, to give a radiant power per unit wavelength, $P_{tel}(\lambda) = A_{tel}F_{tel}(\lambda)$. It subsequently passes through or is reflected off a series of optical surfaces, e.g. the primary mirror, folding mirror, pickoff mirror, dichroics etc.  These are termed the `common optics'.  
 On encountering each optical surface, the stellar spectrum is attenuated by a transmission or reflectance factor, i.e. throughput. The power per unit wavelength after passing through the common optics is 
$P_{ch}(\lambda) = P_{tel}(\lambda)\eta_{tel}(\lambda)$, where $\eta_{tel}(\lambda)$ is the total throughput of the common optics.  The wavelength-dependent throughput of each optical surface is defined in a reference file for the specific instrument.
This constitutes the telescope model and is implemented in the \textit{Instrument} module.   The module then continues the modulation of the signal as follows.

\subsubsection{Channel}

From the common optics, light is passed into one or more instrument channels, which are specialised optical systems designed to examine a specific aspect of the signal.  ExoSim allocates a \textit{Channel} object class to each instrument channel.  Each channel has its own set of optical surfaces, which further attenuates the signal, with its own optical prescription.  
An instrument channel can be a spectrometer or photometer. 
Light may enter the spectrometer through a slit (to reduce background and stray light), or it may be slitless. It is then collimated, dispersed (by a grism, grating or prism) and then focussed onto the detector by camera optics.  Different instrument channels may have different wavelength coverages, different spectral resolving ($R$) power for the dispersion element, and different detector characteristics. Here we do not discuss Fourier transform or integral field spectrometers as currently ExoSim does not simulate these.
Throughput files are used for each channel optical surface.  The power per unit wavelength reaching the detector is then $P_{det}(\lambda) = P_{ch}(\lambda)\eta_{ch}(\lambda)$, where $\eta_{ch}(\lambda)$ is the total throughput of the channel optics.

\subsubsection{Detector}

\begin{figure}
\begin{center}
\includegraphics[width=1\textwidth]{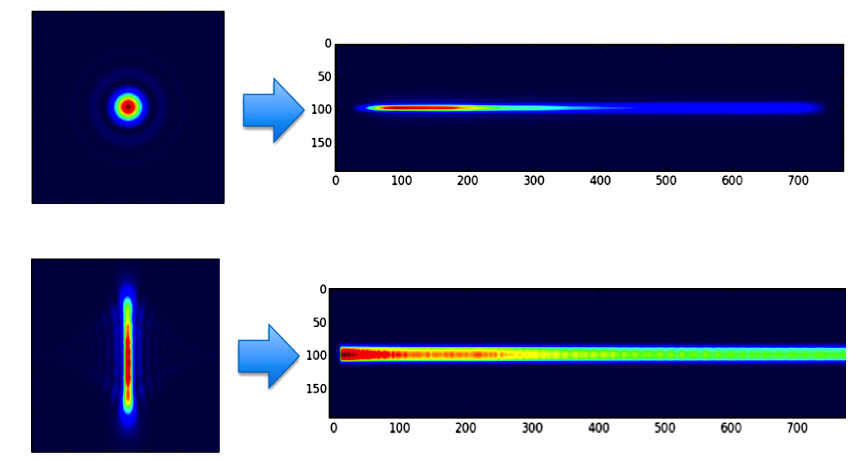}
\caption[]{Spectral images in ExoSim, produced from the co-addition of 2-D PSFs. Top: Spectral image constructed from Airy function PSFs.  Bottom:  Spectral image constructed from an aberrated PSF generated from third-party software; with a few samples provided at different wavelengths, the intervening PSFs can be constructed by interpolation.}
\label{Chapter2:Figure:psf}
\end{center}
\end{figure}

Before falling on the detector, the light has been convolved with the 2-D optical PSF (which is a function of wavelength), and in the case of a spectrum, a spectral trace is projected onto the detector.  These factors determine the spatial distribution of photons on the detector array, i.e. the spectral image. The `wavelength solution' for each pixel $\lambda(x)$, where $x$ is the pixel coordinate in the spectral direction, is obtained as an external input from optical modelling of the dispersion element.
For each pixel column $x$, the corresponding value of $\lambda(x)$ is used to generate a 2-D PSF. The volume of the PSF equals the power falling over the wavelength span of the pixel column.  The PSF convolution is completed by the coadding of these 2-D PSF images onto a 2-D array representing the detector pixel array. Each 2-D PSF is centered on a pixel $x,y$, where $x$ is the corresponding pixel column, and $y$ is the pixel coordinate in the spatial direction and is assumed to be constant.  A spectral image is thus produced as shown in Figure \ref{Chapter2:Figure:psf}.  As shown, PSFs can not only be constructed from Airy functions, but also abberated PSFs provided by third-party software or from real instruments can be used.  This is an improvement over EChOSim which used only Gaussian functions for the PSF.
ExoSim can utilise non-linear wavelength solutions (e.g. from a prism) in additional to linear solutions (from grisms or gratings), whereas EChOSim modeled only linear solutions.   The  convolution with the PSF is performed in 2-D in ExoSim, rather than just 1-D in EChOSim\footnote{In practice, the PSF convolution is performed on a sub-pixelised grid, which ensures Nyquist sampling of the PSF, however for clarity we describe the algorithm using whole pixels.}.

\begin{figure}
\begin{center}
\includegraphics[width=0.8\textwidth]{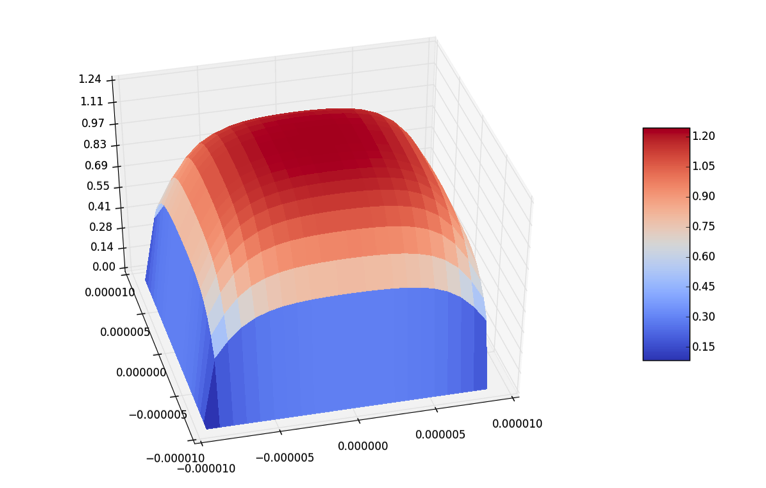}
\caption[]{2-D pixel response function used in ExoSim to simulate intra-pixel variation in responsivity.
x and y axes are show distance in units of m; z axis shows the responsivity after normalizing the volume to unity.
}
\label{Chapter2:Figure:prf}
\end{center}
\end{figure}

As a result of the PSF convolution and spectral dispersion, each pixel will have a radiant power, $P_{pix}(x,y)$.  The detector is a 2-D pixel array, with a given quantum efficiency, $QE(x,y)$, which affects the conversion rate of incident photons to electrons.  
The number of photoelectrons per second produced per pixel is then: $Q(x,y) = P_{pix}(x,y)QE(x,y))hc/\lambda(x)$, where $h$ and $c$ are the Planck constant and speed of light respectively, and $\lambda(x)$ is the central wavelength of the pixel column, $x$.
The final electron count rate will also depend on the intra-pixel response function.  We take the 1-D intra-pixel response function described in Eq. 12 of \cite{Pascale2015} (which models a gradual fall in responsivity with distance from the centre of the pixel) and generalise this to 2-D (example shown in Figure \ref{Chapter2:Figure:prf}).  After normalising this to a volume of unity, we convolve it with the detector array, such that each point on the convolved 2-D array gives the electron count rate over a pixel-sized area, and incorporates the effect of the intra-pixel response function.  Downsampling to whole pixel positions then gives the count rates for each pixel on the focal plane detector array.

Compared to EChoSim, ExoSim can model both photometers as well as spectrometers. In a photometer there is no dispersing element, and an image of the star is formed from all wavelengths over a given bandpass.  A select number of wavelength-dependent 2-D PSFs are generated 
covering the wavelength band of the photometer, which are then  coadded to the 2-D detector pixel array over the same location to produce the final photometric image.

Another source of photons is the thermal emission from telescope and channel optical surfaces.  Like zodiacal light this is a diffuse source.  ExoSim uses essentially the same algorithm for diffuse radiation as explained in \cite{Pascale2015}.

\subsection{Timeline generator}

An observation will attempt to follow the entire transit (or eclipse) event, usually including a similar duration of time out- of-transit.  The observation is divided into a series of exposures, between which the detector array is reset.  Within each exposure the pixels will accumulate electrons as a `ramp'. Each exposure consists of non-destructive reads (NDRs) where snapshots of the accumulating count are read at various times up the ramp.  In data reduction, an image per exposure is obtained, either through fitting the NDR ramp gradient or through last-minus-first processing (correlated double sampling), where the first NDR is subtracted from the last. Up-the-ramp fitting reduces read noise, while correlated-double-sampling removes reset noise.   There may be `dead time' within an exposure cycle, due to reset time and detector idling, making it less than 100\% efficient.  

The \textit{Timeline generator} module allows for dedicated control of time domain elements: the observational timeline, the timing of exposures, and the timing of exposure cycle elements, such as NDRs and detector `dead' times.  The focal plane detector array generated in \textit{Instrument} is used to setup an 3-D array of NDR images against time.  Multiplying by the integration time for each NDR, the pixel photoelectron counts per NDR can be generated. The time-dependent transit light curves are then applied to these counts\footnote{For algorithmic reasons these are in practice applied within the \textit{Noise} module}. Application of the transit light curve can be omitted if a simple out-of-transit simulation is needed.  

Time-dependent instrumental systematics or astrophysical processes such as star spots, can bias or distort the light curve, and therefore impact on the signal variations captured in the NDR time series.  The \textit{Timeline generator} module allows ExoSim to interface with external time domain simulators that model such processes. Timelines of modulated wavelength-dependent variations in the signal from various processes can be produced from these external models, and then applied to ExoSim light curves in this module.  An example are the effects from stellar pulsation and granulation \citep{Sarkar2018}.  Another example is the effect of star spots and faculae on the light curve,  We have developed a dedicated star spot simulator called `SpotSim', described further below.

\subsection{Noise}

\begin{figure}
\begin{center}
\includegraphics[width=0.8\textwidth]{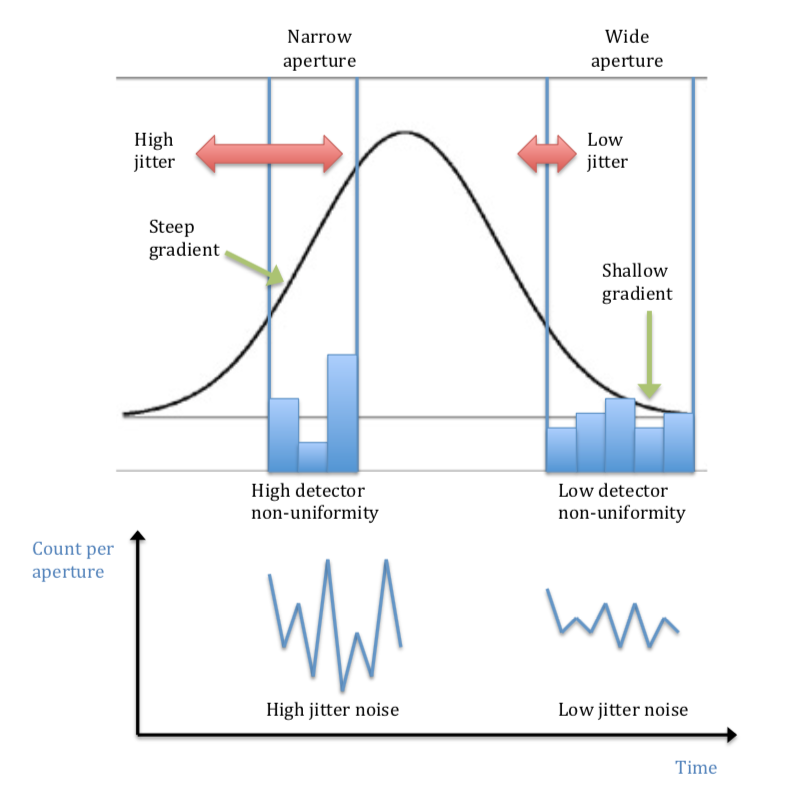}
\caption[]{Factors affecting the noise from pointing jitter. In this simplified case, a 1-D signal (black line) is jittering over a 1-D detector array with pixels of varying quantum efficiency. A narrower aperture, larger QE variation, larger signal gradient and higher pointing jitter (left) results in more noise compared to a wider aperture, smaller QE variation, smaller signal gradient and lower pointing jitter (right).}
\label{Chapter2:Figure:jitter_mech}
\end{center}
\end{figure}

All observations will be subject to random statistical noise. These include Poisson noise processes, also known as `shot' noise. Possion noise is generated from the target itself, as well as from the diffuse sources and from the dark current within each detector pixel.  Read noise occurs from the uncertainty in the conversion of charge in the on-chip amplifier to analogue voltage.  It can be modelled as a Gaussian distribution around the final pixel count.  The above noise types are `white noise' processes, where there is no correlation between different reads.  Hence ExoSim applies these noise effects in a \textit{Noise} module where the individual pixel counts per NDR are randomly adjusted to simulate the effect of these noise processes.  For correlated noise, there is stastistical dependence between different measurements, and thus a different approach is taken when simulating these.  If the correlated noise is wavelength-dependent (but not dependent on spatial position), one way is to apply the effects of the noise on the light curves in the \textit{Timeline generator} module.  The noise timelines are generated by an external model of the correlated noise process. This approach was used to apply correlated noise from stellar pulsation and granulation in \cite{Sarkar2018}.  

Pointing jitter is a type of instrumental correlated noise that is complex in origin, depending on several factors: the power spectrum of the jitter, integration time, source brightness, intra- and inter-pixel responsivity variations and the application of apertures or bins (Figure \ref{Chapter2:Figure:jitter_mech}).  
ExoSim incorporates a jitter simulation subroutine, which improves on that in EChOSim in the following ways.  The EChOSim jitter simulation only considers spatial jitter noise, whereas ExoSim treats jitter in 2-D modeling both spectral and spatial jitter simulataneously.  EChOSim used a parametric model prediction for the jitter noise, whereas in ExoSim, jitter is simulated in a dynamical model which more accurately represents the physical process.

Pointing jitter timelines are generated for $x$ and $y$ directions from a power spectral density (PSD) frequency spectrum. The PSD is randomised in amplitude and phase so that random jitter timelines for each axis are produced after inverse Fourier transformation.  The time step duration in the jitter timeline is calculated to ensure Nyquist sampling of the jitter PSD. The code iterates through all jitter timesteps, sampling the convolved focal plane detector array over each whole pixel position, but with slightly different offset positions in $x$ and $y$ each time; the sampled positions thus give the count rate over whole pixels at that jittered position.  If required, the convolved focal plane detector array is first upsampled to ensure that the jitter rms is Nyquist sampled during this process. For each NDR, the cumulated counts per jittered timestep are co-added to give a final NDR count.  An inter-pixel variation in QE is modeled (typically 5\% rms), as Gaussian distribution upon the baseline QE of the detector; this is an important factor in producing jitter noise, and is applied to the NDR counts.
Therefore, unlike EChOSim which acccounted only for the photometric variation from jitter between integrations, ExoSim in addition captures the effects of jitter within the integration time.  This manifests as blurring of the image. 
 
\begin{figure}
\begin{center}
\includegraphics[width=1.0\textwidth]{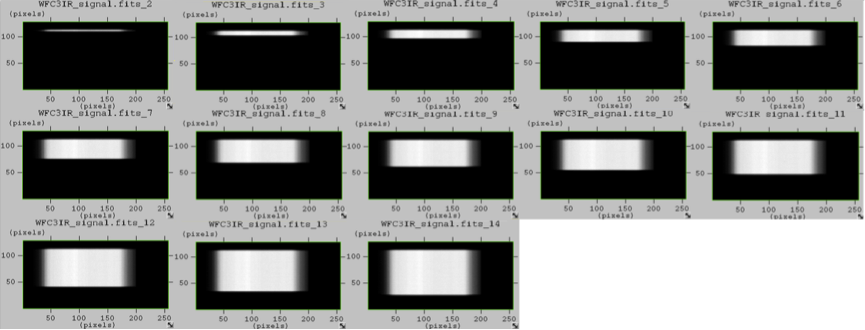}
\caption[]{Example of simulated scanning mode in ExoSim.  ExoSim models the Hubble WFC3 IR instrument observing the super-Earth GJ 1214b. This exposure consists of 13 NDRs. The progression of the scan is evident in the sequence of NDR images.}
\label{Chapter7:Figure:scan}
\end{center}
\end{figure}

Another innovation in ExoSim is the ability to simulate scanning mode observations (Figure \ref{Chapter7:Figure:scan}), as performed on the Hubble WFC3 IR instrument for exoplanet transit spectroscopy \citep[e.g.][]{Kreidberg2014}.  This is implmented by applying a sawtooth profile to the $y$ jitter timeline.

\subsection{Output}

ExoSim uses the same FITS file format for its output as EChOSim.  In ExoSim, however the output files will contain multiple NDRs per exposure, which is closer to the usual raw image output of an instrument.  EChOSim in contrast generated images per exposure, that assumed post-processing had been performed, e.g read noise was reduced assuming up-the-ramp processing but without the production of actual ramps.  Thus the output from ExoSim provides more realistic mock data for the validation of data reduction pipelines.

\section{SpotSim}

\begin{figure}
\begin{center}
\captionsetup{justification=centering}
  \begin{subfigure}[b]{0.49\textwidth}
    \includegraphics[trim={0 0 0 0}, clip,width=\textwidth]{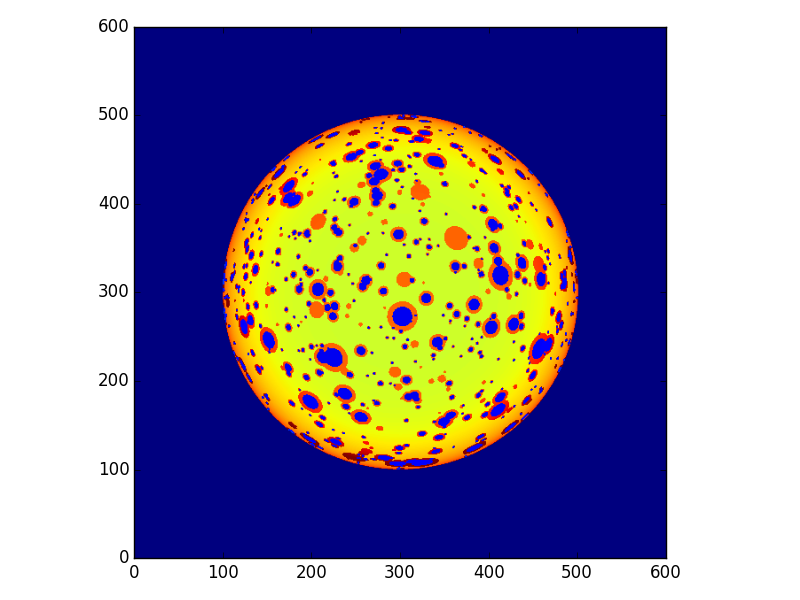}
    \caption{ Uniform random spatial distribution\\Log-normal size distribution}
    \label{fig:1}
  \end{subfigure}
  \begin{subfigure}[b]{0.49\textwidth}
    \includegraphics[trim={0 0 0 0}, clip,width=\textwidth]{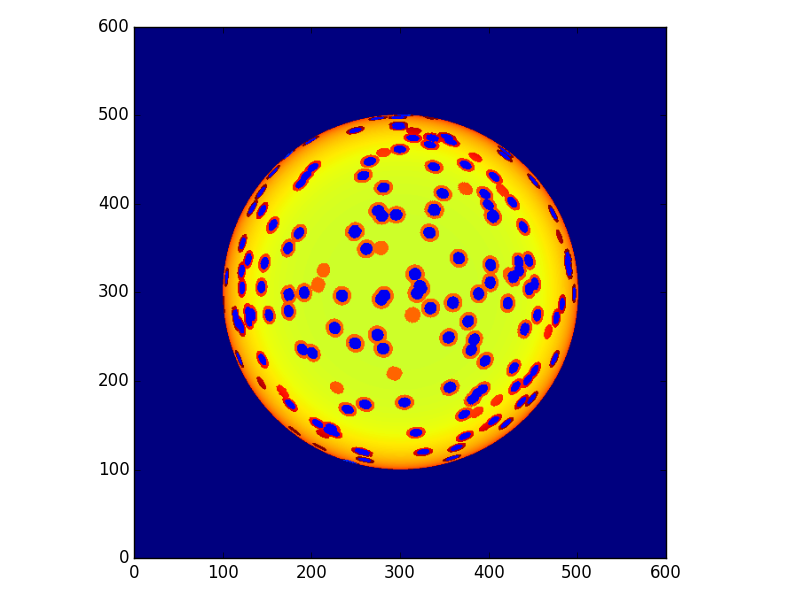}
    \caption{ Uniform random spatial distribution\\Fixed size distribution}
    \label{fig:1}
  \end{subfigure}
  \begin{subfigure}[b]{0.49\textwidth}
    \includegraphics[trim={0 0 0 0}, clip,width=\textwidth]{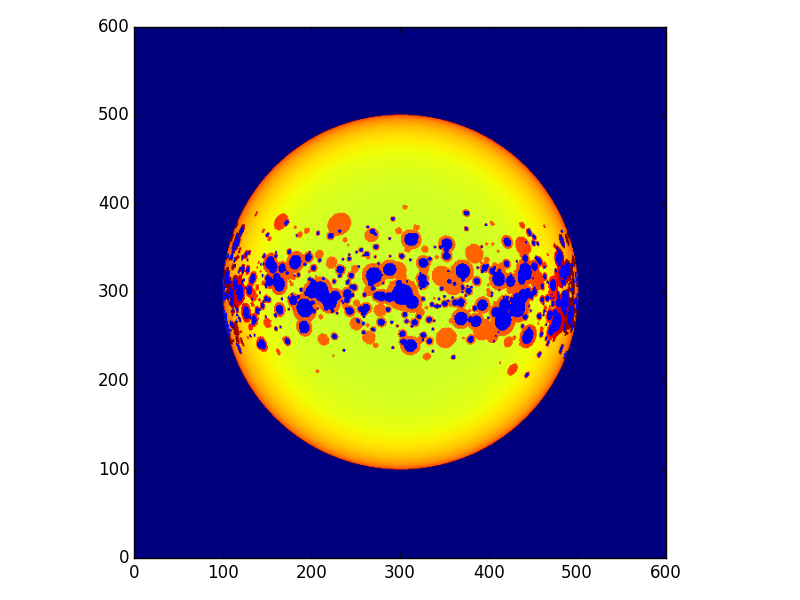}
    \caption{ Latitudinal spatial distribution}
    \label{fig:1}
  \end{subfigure}
  \begin{subfigure}[b]{0.49\textwidth}
    \includegraphics[trim={0 0 0 0}, clip,width=\textwidth]{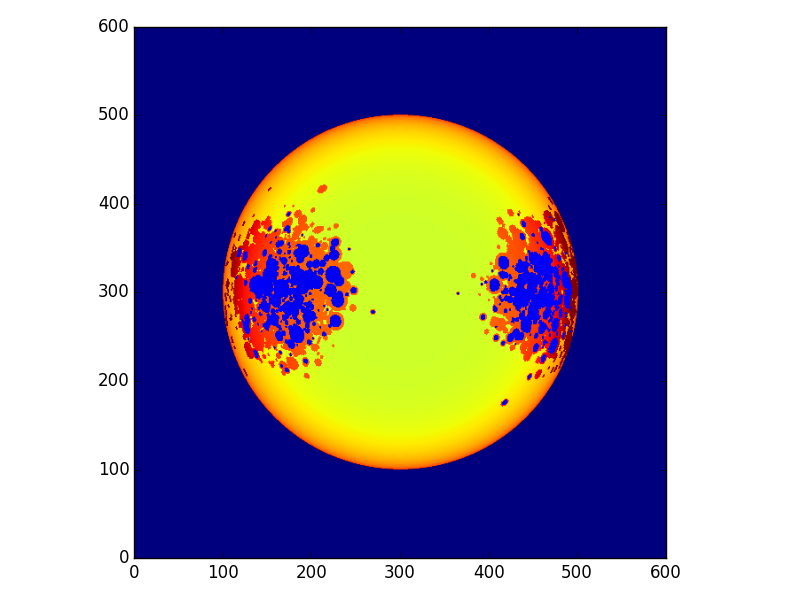}
    \caption{ Longitudinal grouping spatial distribution}
    \label{fig:1}
  \end{subfigure}
\caption[]{Example stellar surface simulations from ExoSim's dedicated star spot simulator, `SpotSim'. Limb darkening is included, and both spots and faculae can be modelled. Spots are shown as blue areas, and faculae as red. Different spatial and size distribution cases are shown.}
\label{Fig:spots}
\end{center}
\end{figure}

\begin{figure}
\begin{center}

\includegraphics[trim={0 0 0cm 0}, clip, width=0.49\textwidth]{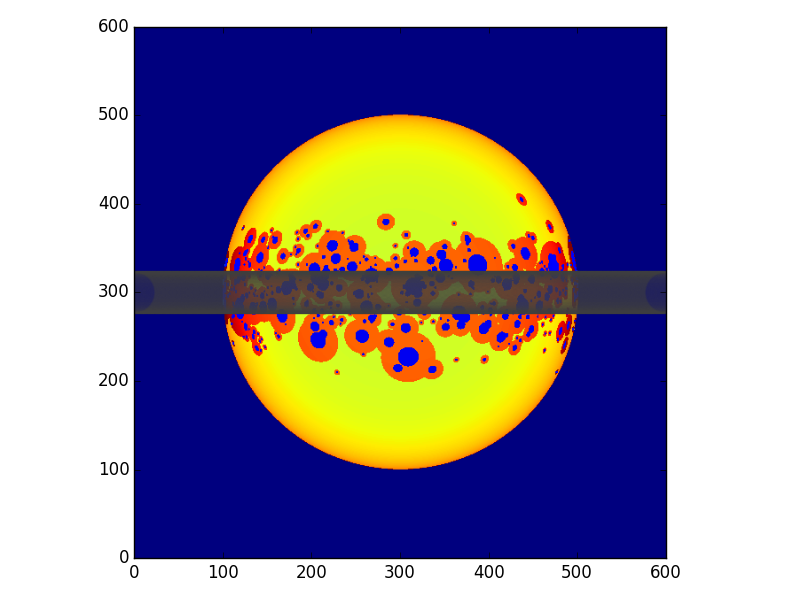}
\includegraphics[trim={0 0 0cm 0}, clip, width=0.49\textwidth]{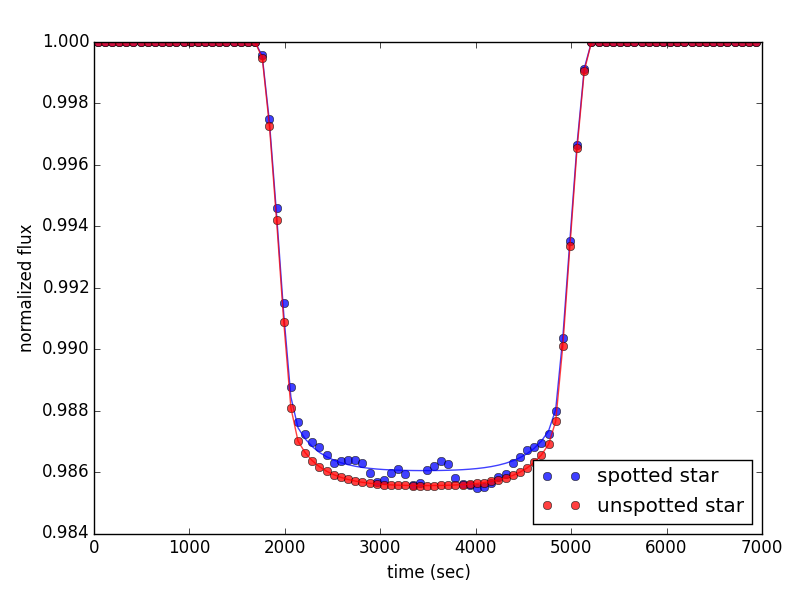}
\caption[]{Example of SpotSim planet transit chord (left) with resulting effect on an ExoSim-simulated spectral light curve at 2.9 \textmu m (right) due to star spot and faculae occultations. In the right figure, red dots show the light curve in the unspotted case, and blue dots in the spotted case. The blue line is a model curve fit to the spotted light curve, showing an underestimation of the true transit depth.}
\label{Fig:spots2}
\end{center}
\end{figure}

A dedicated star spot simulator has been developed for use with ExoSim: `SpotSim'.  SpotSim interacts with ExoSim in the \textit{Timeline generator} module, where it modulates the light curves produced in ExoSim.  SpotSim models a spotted star surface, complete with wavelength-dependent limb darkening.  A transiting planet is simulated, with the planet/star radius and orbital parameters obtained via the ExoSim inputs. The planet transit light curve is obtained by moving the planet in small time steps across the stellar surface.  The resulting light curve will capture the effects of both occulted and unocculted star spots.  In the \textit{Timeline generator} module these SpotSim light curves are applied as corrections to  the unspotted ExoSim light curves, which therefore incorporate these effects, and are used downstream in the remaining ExoSim simulation. SpotSim can also model faculae, which have opposing effects to spots.  The unspotted photosphere, spots and faculae, have different brightness spectra respectively, with spots being colder than the photosphere, and faculae being hotter. The modeling of faculae currently does not include limb distance brightness dependency \citep{Norris2017}. The spatial distribution of spots can be modelled as random uniform, clustered as a latitudinal band, or clustered into longitudinal groupings (Figure \ref{Fig:spots}).  The size distribution of the spots can be modelled as a log-normal distribution \citep{Solanki} (Figure \ref{Fig:spots} A), or as fixed sizes (Figure \ref{Fig:spots} B). The effects of spots and faculae will
be to distort the transit light curve (Figure \ref{Fig:spots2}), and potentially bias the final recovered transit depth. This can manifest on the final reconstructed spectrum as a wavelength-dependent bias, the level of which will depend on many factors, such as size of spots and faculae, the balance between occulted and unocculted features, the spot filling factor, the stellar class, relation of the transit chord to the spatial distribution, etc.  The effects of spots and faculae are thus complex and their impact depends on many factors.  A simulator like ExoSim with SpotSim can thus help to understand the impact of spots and faculae on recovered planet spectra under different conditions.  The ExoSim data output, which will contain spot-contaminated time series spectra, can also be used as test data for spot correction algorithms in data pipelines.

\section{Validation}
\label{sec:2}

\subsection{Validation of focal plane signal}
\label{Chapter2:subsection:vs analytical}

\begin{figure}
\begin{center}
\includegraphics[width=1\textwidth]{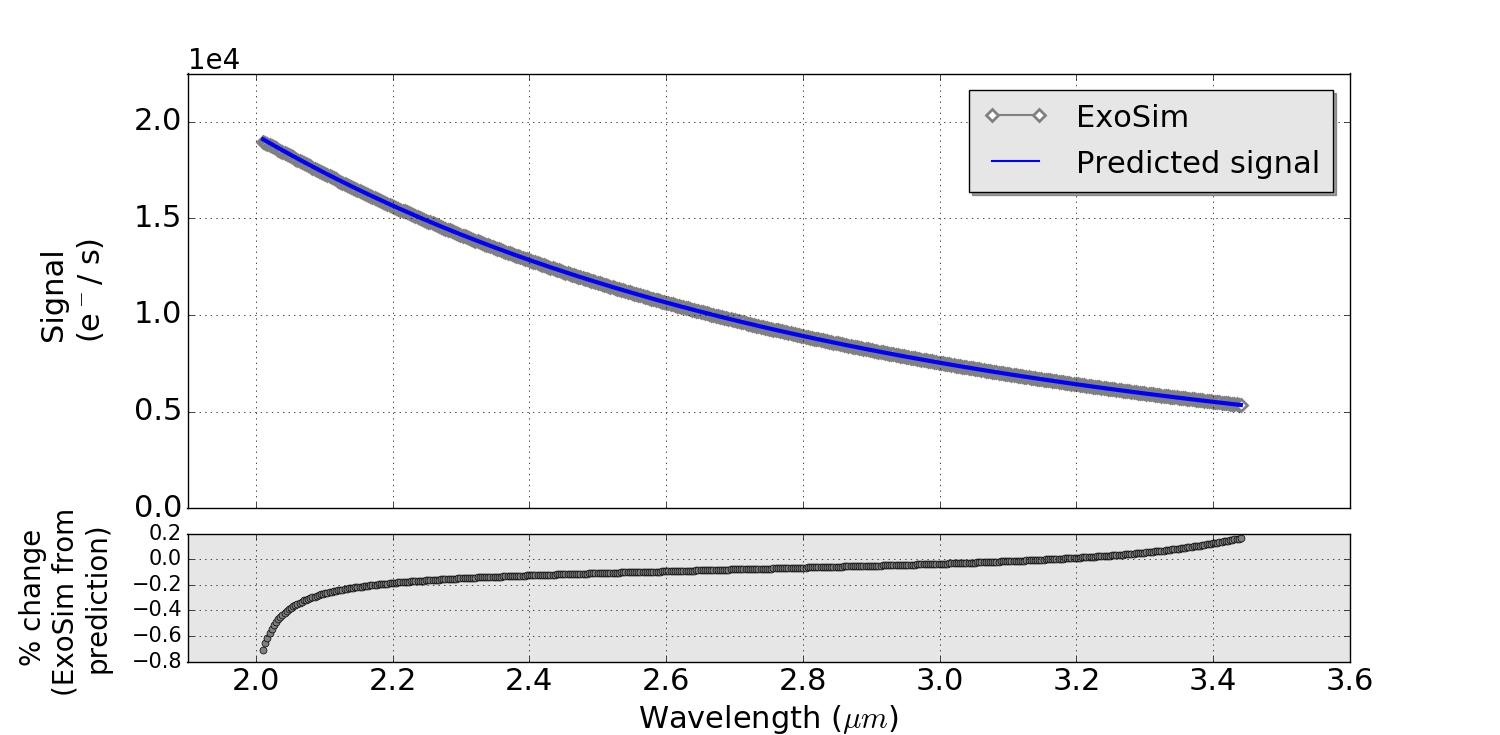}
\caption[]{ ExoSim focal plane signal compared to analytical 
prediction, where the x-axis shows the wavelength on a pixel column. Grey plot shows  percent change of ExoSim from prediction over the wavelength range (2.01-3.44 \textmu m).}
\label{Chapter2:Figure:validation1}
\end{center}
\end{figure}

ExoSim simulations were performed without noise for an out-of-transit  observation of the star 55 Cancri.  The star was modeled using a Planck black body spectrum (T=5196 K) rather than using a PHOENIX spectrum, since this permitted an easier comparison with the validating equation  which also uses a black body function.  The instrument model included a 0.5 m primary mirror telescope with an R = 254, $f$-number = 18.5, IR grating spectrometer. 
From the focal plane image produced in the \textit{Instrument} module, the summed photoelectron count rates per pixel column, $x (\lambda)$,  were obtained. The signal is expected to agree with the following equation:  

\begin{equation}
Q(x) = \pi B_{\lambda(x)}(5196 K)\left( \frac{R_s}{D} \right)^2A_{tel} \eta(x) QE(x)\Delta\lambda(x) \frac{\lambda(x)}{hc}
\end{equation}
\label{Eq: BB}

\noindent where $Q(x)$ is the photoelectron count rate in pixel column $x(\lambda)$,  $B_{\lambda(x)}(5196 K)$ is the Planck function of the star, $\eta(x)$ is the total optical transmission, and $\Delta \lambda(x)$ is the wavelength interval over the pixel column width. The comparison between the ExoSim signal and the prediction from Eq.\ref{Eq: BB} is shown in Figure \ref{Chapter2:Figure:validation1} (where the count rates are plotted against the wavelength of the corresponding pixel column).  We find that over the wavelength range examined, the percent variation from the predicted value is within the range -0.8 to +0.2\%, showing good agreement.  This test therefore validates the basic signal production in ExoSim.

\subsection{Validation of uncorrelated noise}

\begin{figure}
\begin{center}
\includegraphics[width=0.49\textwidth]{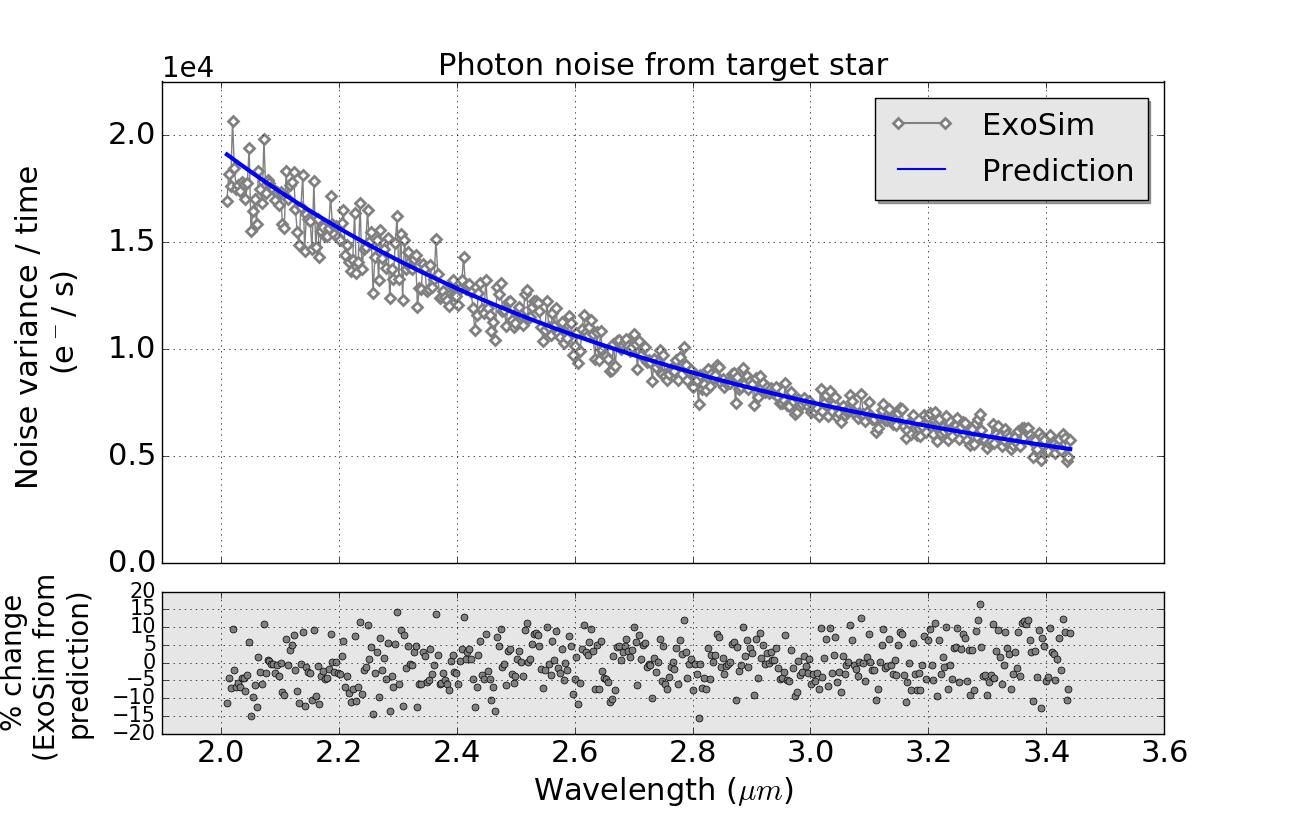}
\includegraphics[width=0.49\textwidth]{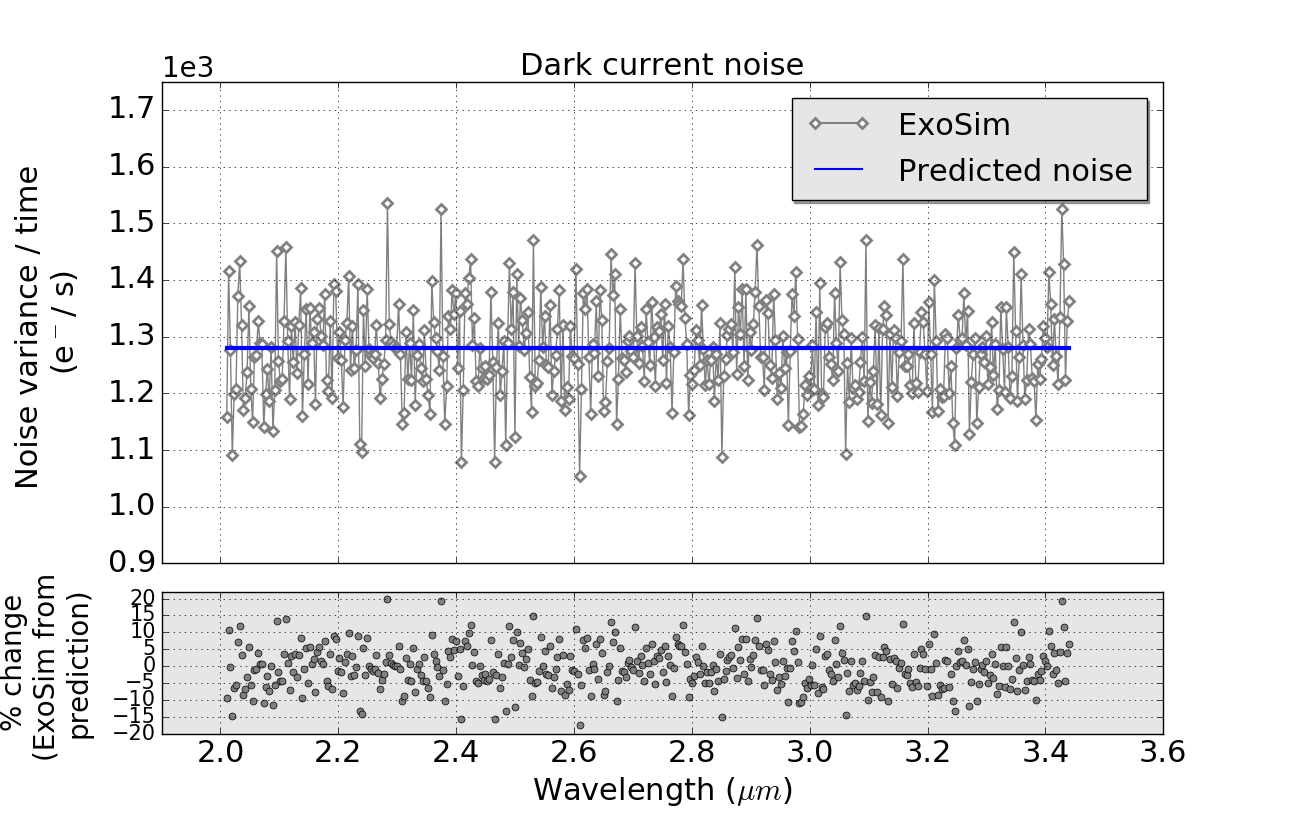}
\includegraphics[width=0.49\textwidth]{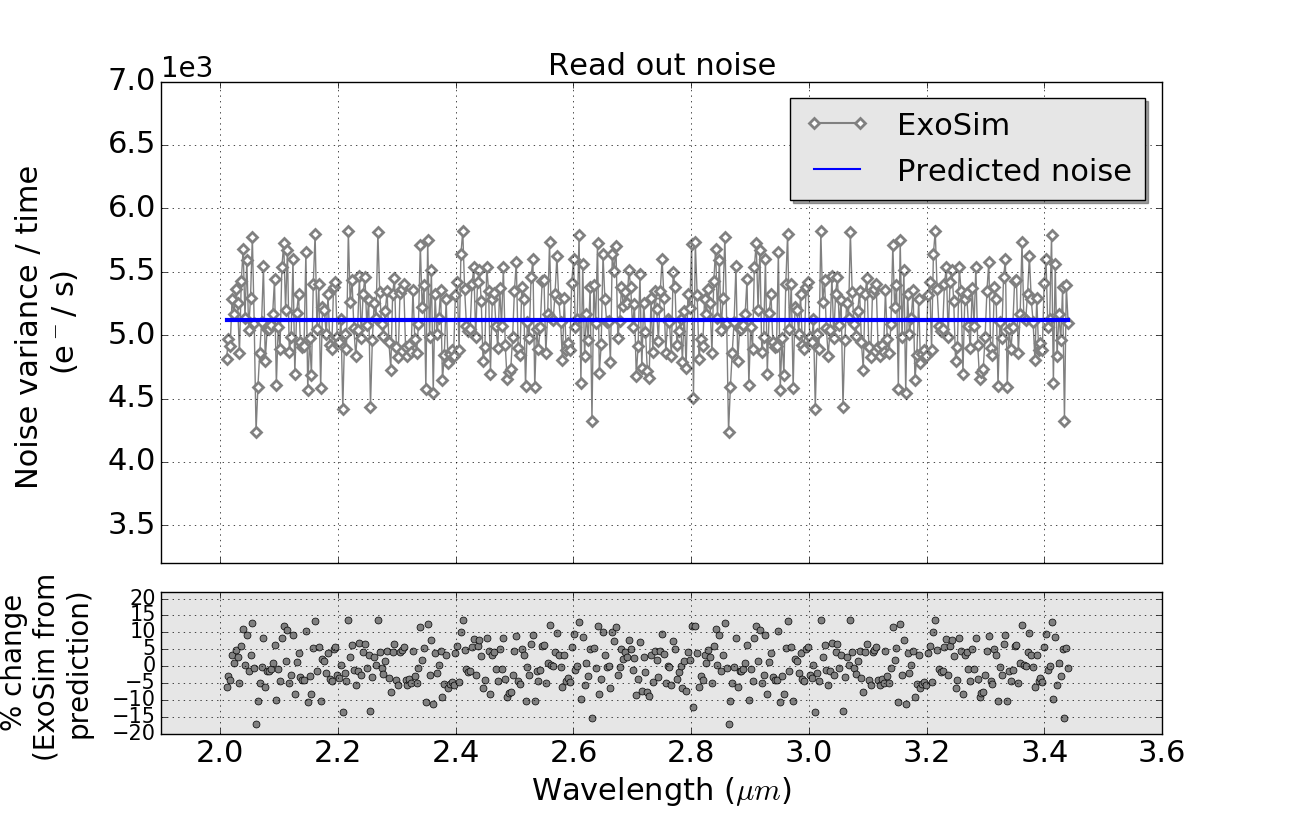}
\includegraphics[width=0.49\textwidth]{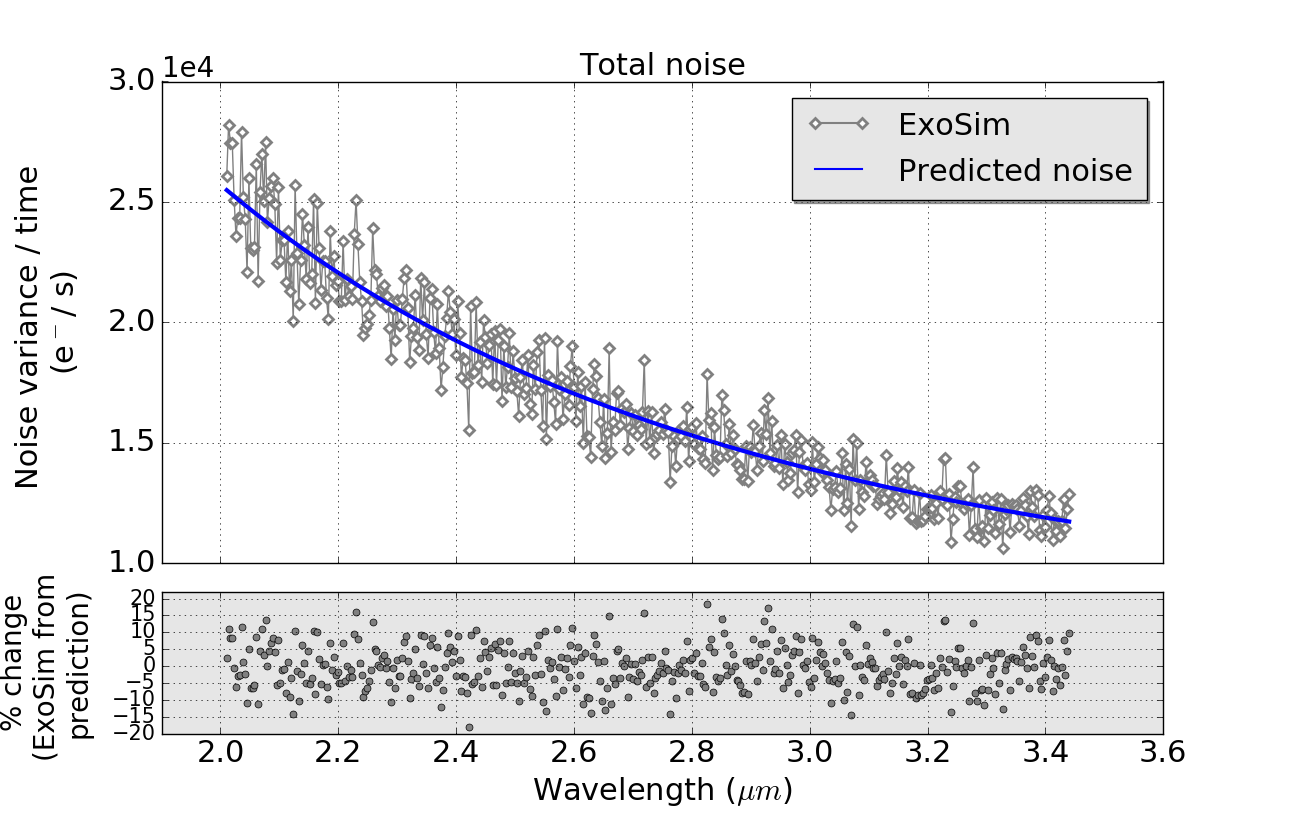}
\caption[]{ExoSim noise sources compared to predicted noise variance per unit time per pixel column, for photon noise from the target star (top left), dark current shot noise (top right), CDS read out noise (bottom left) and combined total noise (bottom right). The x-axis shows the wavelength on the pixel column.  Grey plots show percent difference of ExoSim from the prediction.}
\label{Chapter2:Figure:validation2}
\end{center}
\end{figure}

The same instrument and star models as in Section \ref{Chapter2:subsection:vs analytical} were used in ExoSim, and an out-of-transit observation simulated, where the exposure cycle consisted of two NDRs. Following the correlated double sampling (CDS) method, the first NDR is subtracted from the final NDR, to give a final CDS exposure.  The noise variance  was obtained for the summed count in each pixel column $x$, and divided by the CDS time to obtain a noise variance per unit time in  e$^-$/s.  The results were obtained for different types of uncorrelated, `white' noise sources in isolation: photon noise from the target star, dark current noise and read out noise, and for all these noises sources combined.  The results were tested against predictions from Equations \ref{Eq: PN} to \ref{Eq: TN}.

\begin{equation}
{\sigma_{photon}}^2(x)/t = Q(x) 
\label{Eq: PN}
\end{equation}

\begin{equation}
{\sigma_{dark}}^2(x)/t = N_{pix}I_{dc}
\label{Eq: DN}
\end{equation}

\begin{equation}
{\sigma_{cds}}^2(x)/t =  (2N_{pix}{\sigma_{ro}}^2)/t
\label{Eq: RN}
\end{equation}

\begin{equation}
{\sigma_{total}}^2(x)/t = [{\sigma_{photon}}^2(x) + {\sigma_{dark}}^2(x) + {\sigma_{cds}}^2(x)]/t
\label{Eq: TN}
\end{equation}  

\noindent where $N_{pix}$ is the number of pixels in the column (in this case 64), and $I_{dc}$ is the dark current on a pixel (in this case 20 e$^-$/s). $\sigma_{photon}$ is the target star photon noise, $\sigma_{dark}$ is the dark current shot noise, $\sigma_{ro}$ is the read out noise for a single pixel read (in this case 20 e$^-$), and $\sigma_{cds}$ is the read out noise for a CDS exposure per pixel column. 

Since ExoSim uses stochastically-generated values in a dynamic simulation, the noise variance measurements reflect these random variations from pixel column to pixel column (Figure \ref{Chapter2:Figure:validation2}). Thus when comparing to the prediction from the equations, we use the mean percentage difference over the wavelength range studied as a measure of similarity.  These are  -0.16, -0.22, +0.15 and -0.22 \% for photon noise, dark current noise, read noise and total noise respectively.  Thus there is good agreement with the equations. This test also showed that these noise sources come from statistically independent processes.

\subsection{Validation of pointing jitter model}

\begin{figure}
\begin{center}
\includegraphics[trim={0 0 0 0}, clip,width=0.7\textwidth]{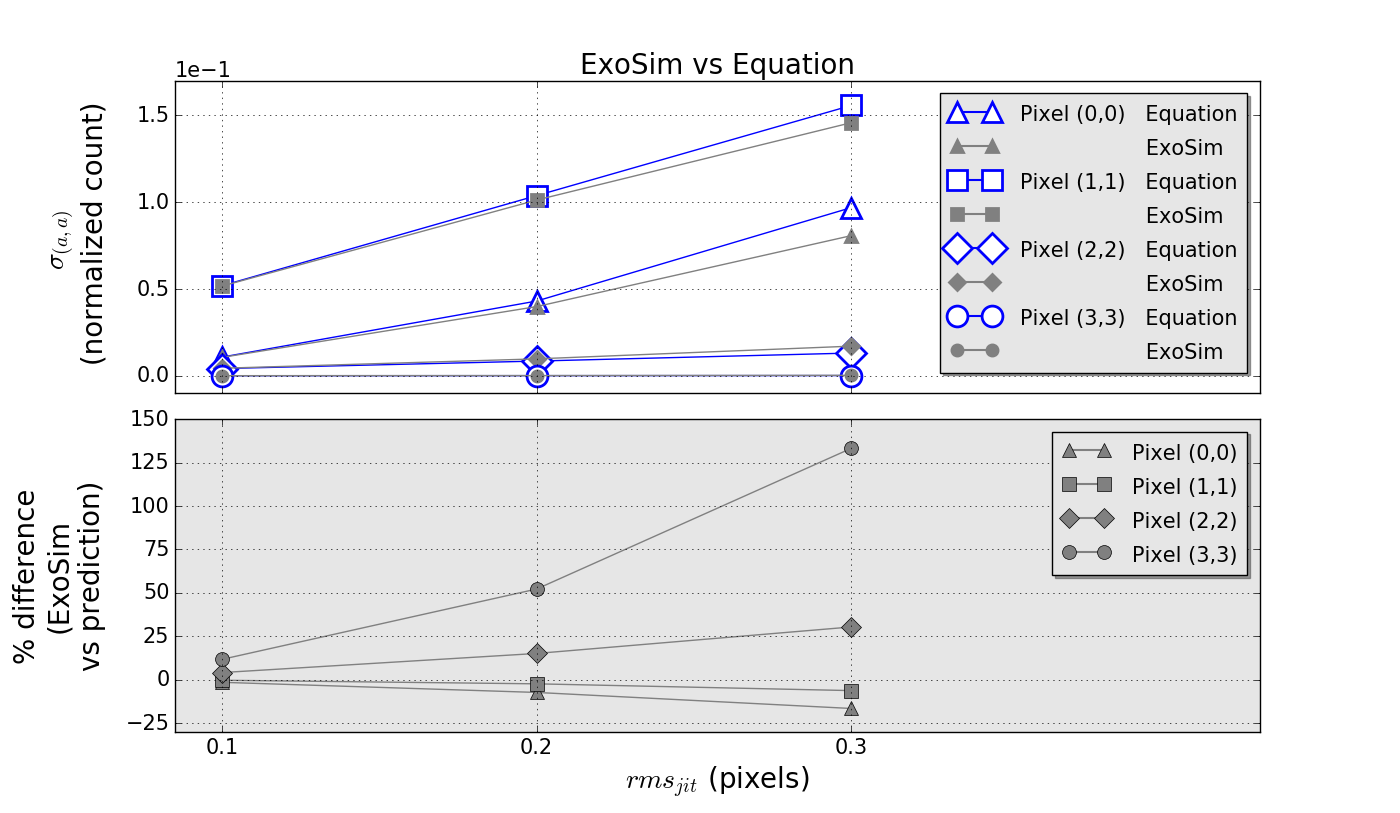}
\includegraphics[width=0.7\textwidth]{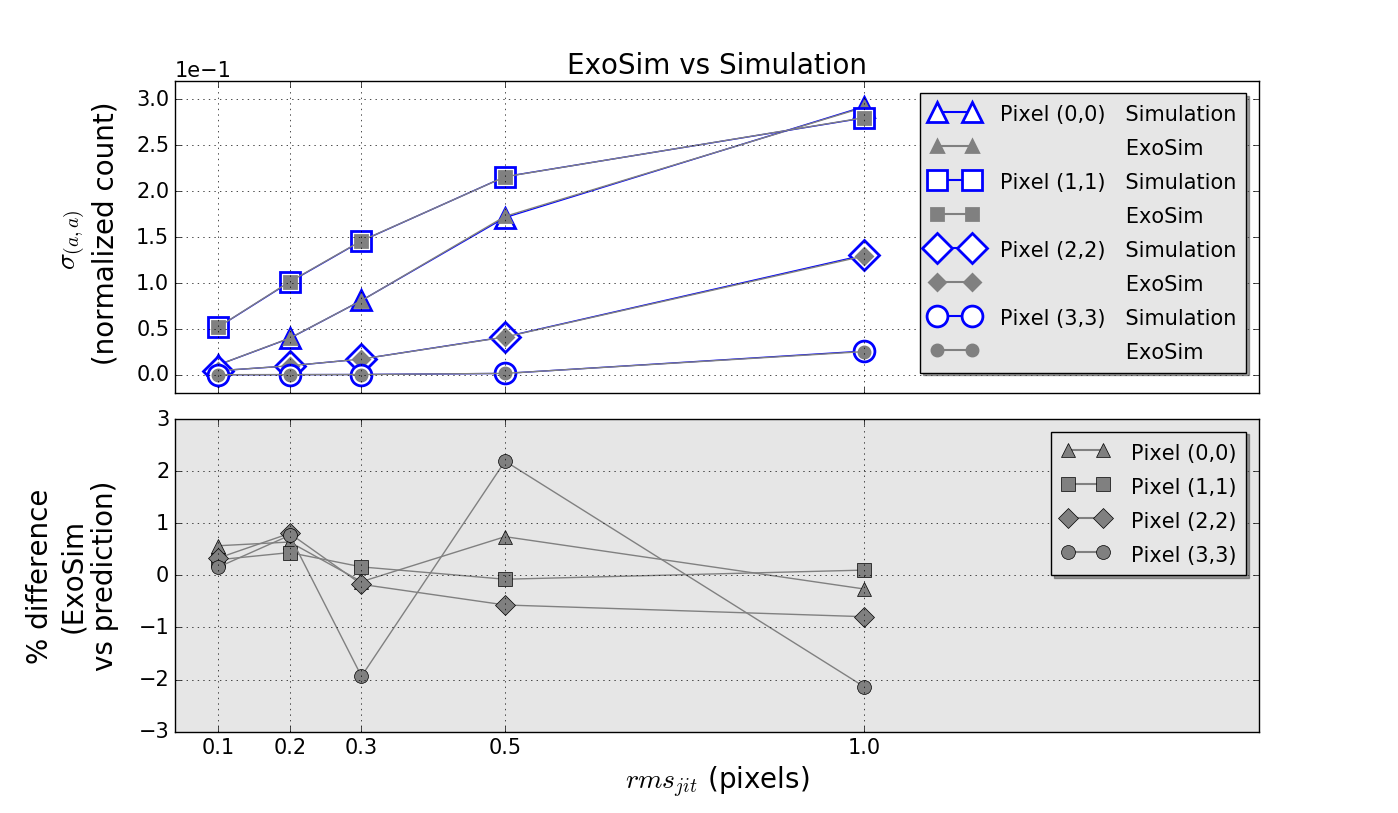}

\caption{ExoSim jitter noise compared to predictions for pixel positions shown. Top: ExoSim compared to Equation \ref{Equation:jitter}. Bottom: Exosim compared to independent simulation.  Grey plots show percentage difference of ExoSim from the prediction.}
\label{Chapter2:Figure:jitter val}
\end{center}
\end{figure}

The mechanism of the jitter code was tested for validity against a prediction from an analytic expression.  To accomplish this, ExoSim was run under a simplified set of conditions: 1) a focal plane with all pixels having an identical, flat response (i.e. no intra- or inter-pixel variations), 2) a source consisting of a monochromatic beam with a Gaussian PSF with full-width-at-half-maximum (FWHM) of 2.27 pixels, 3) jitter with a flat PSD giving a jitter timelines in each axis with standard deviation, $rms_{jit}$.  The beam was centered on a central pixel, given a pixel coordinate (0,0), and allowed to jitter in 2-D around this. The count on this pixel at each jitter time step was monitored and a timeline of counts obtained. 
Pixels at progressively further distances from the central pixel were similarly monitored.  These were pixel (1,1) (i.e. one pixel row above and one pixel column to the right of the central pixel), pixel (2,2) and pixel (3,3).  The measured counts were all normalised to the central pixel count obtained with a stationary beam.
The standard deviation of the normalised counts 
in these pixels resulting from the jittering beam were compared to a prediction from Eq. \ref{Equation:jitter}.  This equation is based on a second order Taylor expansion, and gives the expected variation in the value ('count') of a 2-D Gaussian function (of unity maximum) as the measurement point is jittered in 2-D. 

\begin{equation}
\sigma_{(a,a)} \approx 
\left[
\exp(-a^2/2s^2) \left(4{rms_{jit}}^4\left[\frac{a^2-s^2}{s^4}\right]^2 
+ {rms_{jit}}^2\left[\frac{2a^2}{s^4}\right]
\right)
\right]^{1/2}
\label{Equation:jitter}
\end{equation}
where $\sigma_{(a,a)}$ is the standard deviation of the count predicted at position (a,a), and $s$ is the standard deviation of the 2-D Gaussian ($s \approx$ FWHM/2.355). This expression will be most accurate for small values of $rms_{jit}$.  The predicted values of $\sigma_{(a,a)}$ for a 2-D Gaussian matching the ExoSim beam were calculated and compared to the ExoSim results.  The comparison is shown in Figure \ref{Chapter2:Figure:jitter val}, top.  In this figure, $\sigma_{(a,a)}$ is the standard deviation of the normalised counts in each ExoSim pixel $(a,a)$, and the predicted standard deviation from Eq. \ref{Equation:jitter}.
ExoSim matches the prediction best at lower values of $rms_{jit}$.  If $rms_{jit}$ is 0.1 pixels, ExoSim is within 5\% of the prediction for $a<3$. However the percent deviation worsens with larger values of $rms_{jit}$, as may be expected due to the Taylor approximation, and are worst at $a=3$.

To assess validity of the ExoSim jitter simulation at higher values of $rms_{jit}$, we performed a separate computer simulation of a 2-D Gaussian function, matched in size to the ExoSim beam, with a maximum of unity.  The 2-D Gaussian was sampled randomly $10^6$ times around points ($a,a$) [$a$=0,1,2,3], with a standard deviation of $rms_{jit}$ in each axis.  The resulting standard deviations of the sample values, $\sigma_{(a,a)}$, at each point ($a,a$) were compared with ExoSim results, as shown in Figure \ref{Chapter2:Figure:jitter val}, bottom.
We obtain a good agreement upto the higher values of $a$ and $rms_{jit}$, ExoSim always being within 3\% of the independent simulation. These results verify the accuracy of the baseline mechanism used in ExoSim's pointing simulation.

\subsection{ExoSim vs ESA radiometric model}
\label{Chapter2:subsection:vs ESA}

The European Space Agency Radiometric Model (ERM) was developed independently as simulator for the EChO and ARIEL missions \citep{Tinetti2018}. Unlike ExoSim, the ERM does not perform a dynamical simulation, but a `static' one, using a set of parametric equations to obtain an estimate of signal and noise for transiting exoplanet observations.  Results are rapid which make it ideal for estimating the SNR of a large number of exoplanet observations in a survey, as demonstrated on the ARIEL target list 
\citep{Zingales2018}. A disadvantage of static simulators is that since they do not model the time domain directly, they cannot capture the effects of correlated noise or time-dependent systematics\footnote{In such circumstances a noise margin is usually added to ensure that an SNR estimate from such a simulator is not overly optimistic.}
Obtaining similar results across different simulators for the same conditions adds confidence to the results of both simulators.  
We therefore set up a cross-validation test to see if similar results were obtained between ExoSim and the ERM.

\begin{figure}
\begin{center}
\includegraphics[trim={0 0 2cm 0}, clip, width=0.49\textwidth]{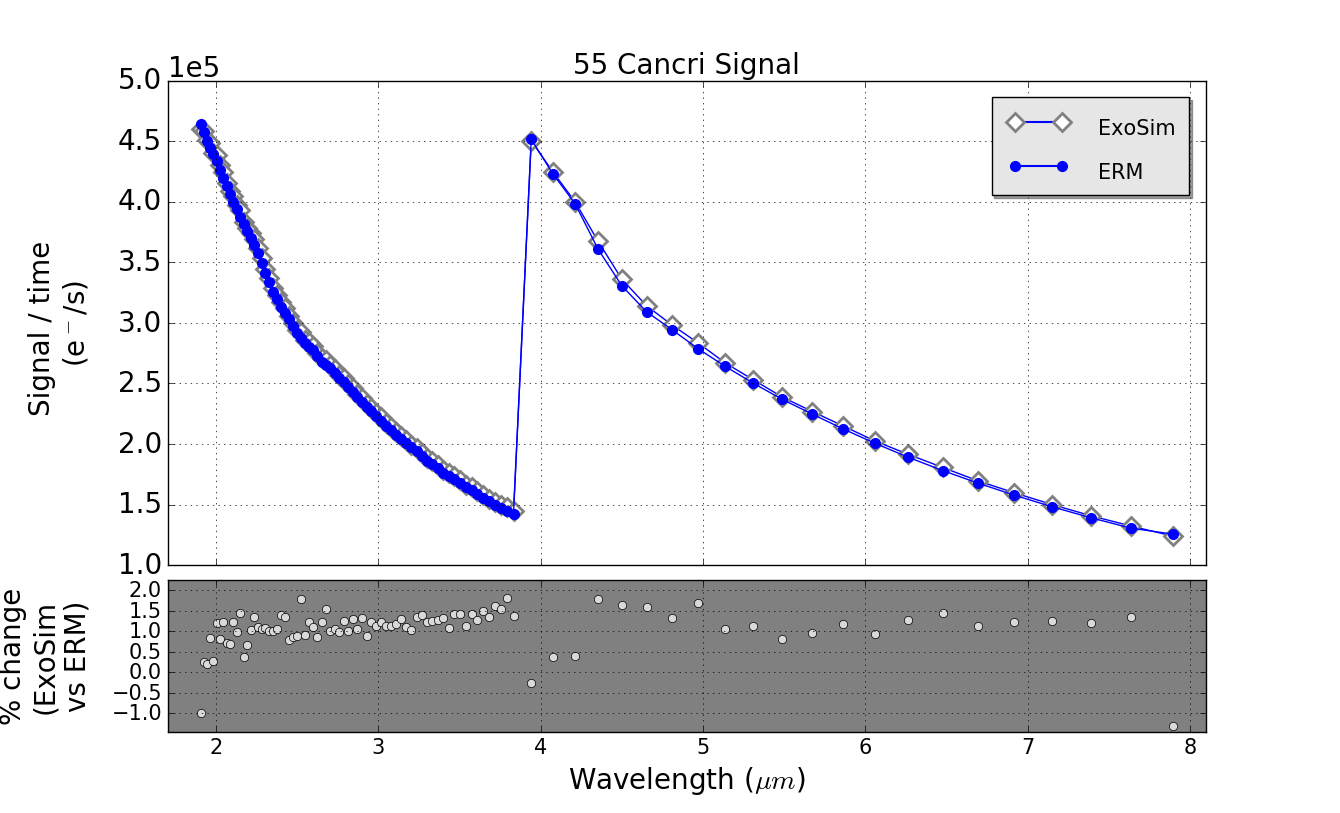}
\includegraphics[trim={0 0 2cm 0}, clip,
width=0.49\textwidth]{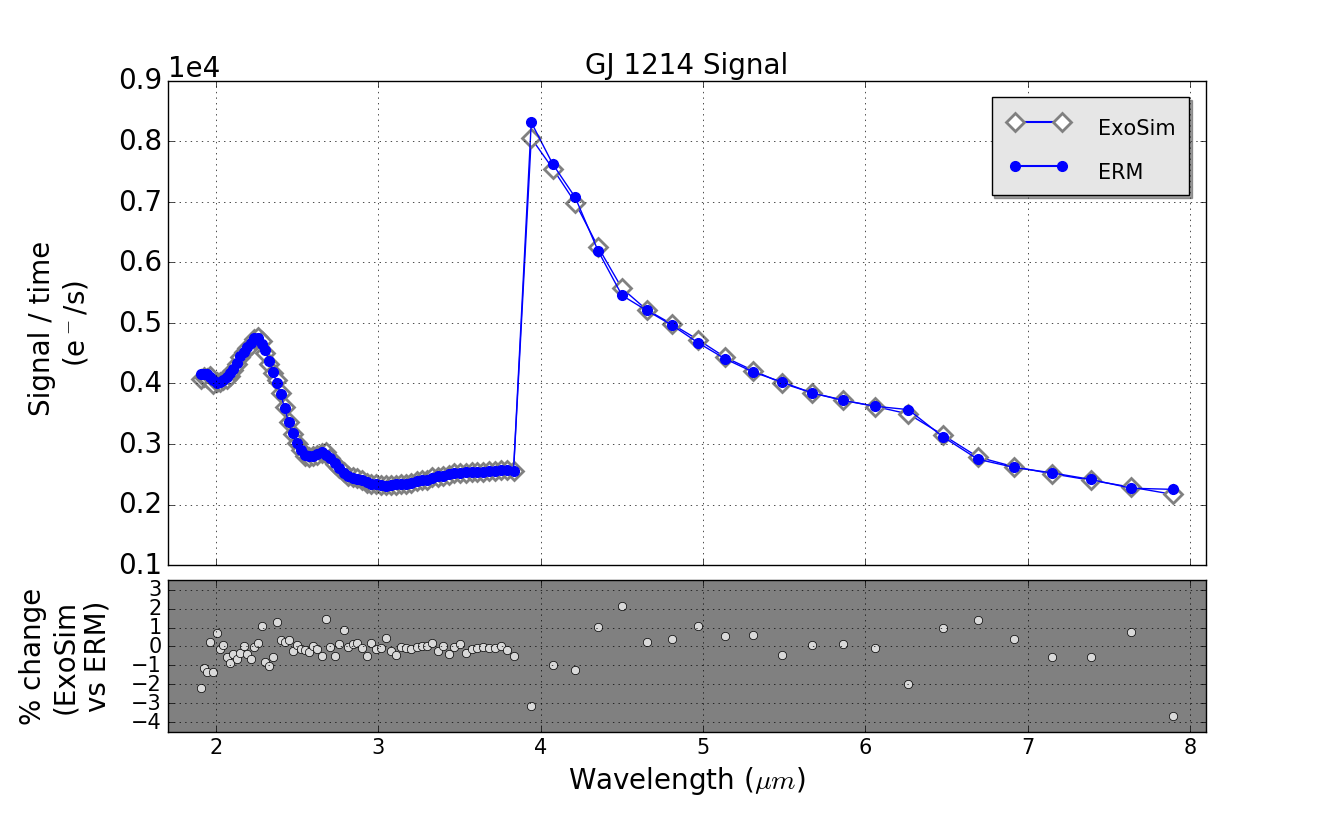}
\includegraphics[trim={0 0 2cm 0}, clip,
width=0.49\textwidth]{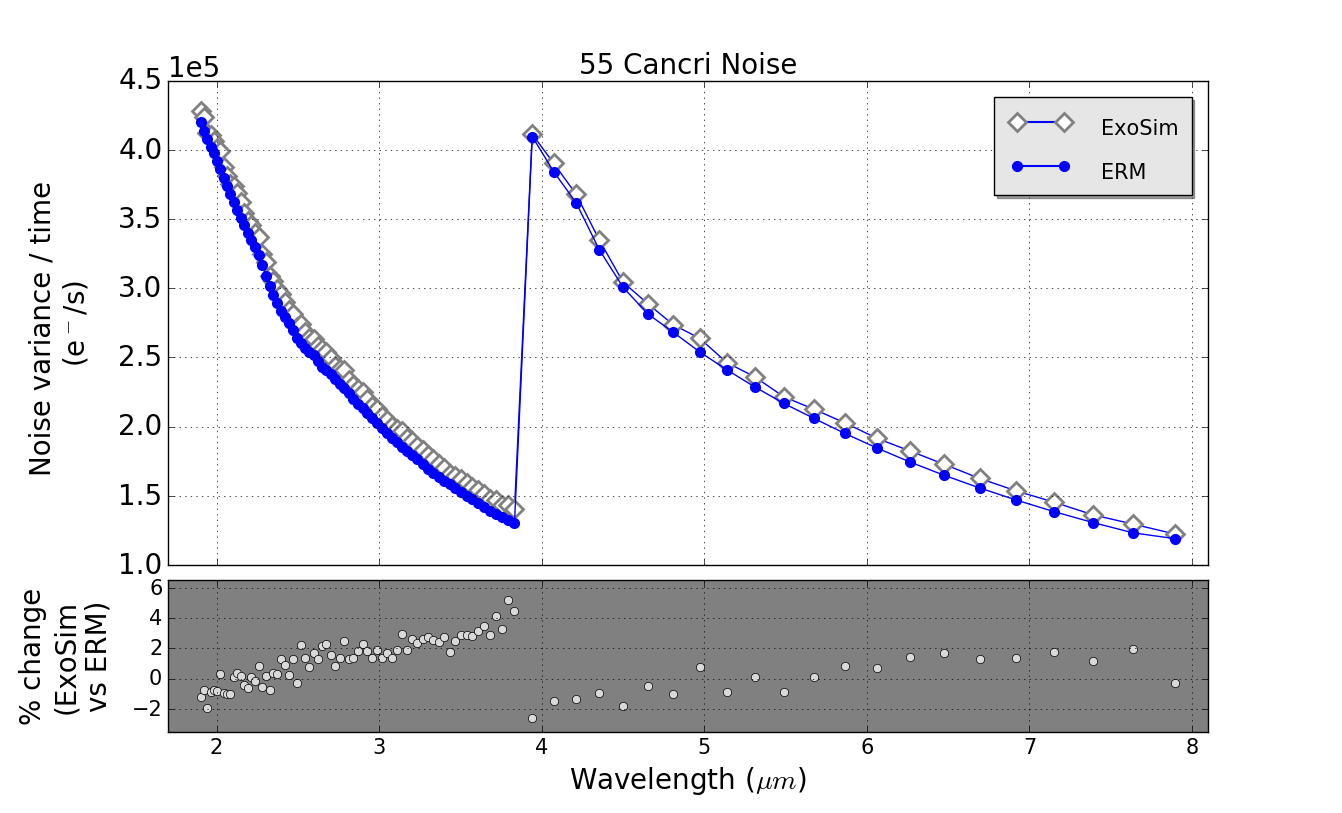}
\includegraphics[trim={0 0 2cm 0}, clip,
width=0.49\textwidth]{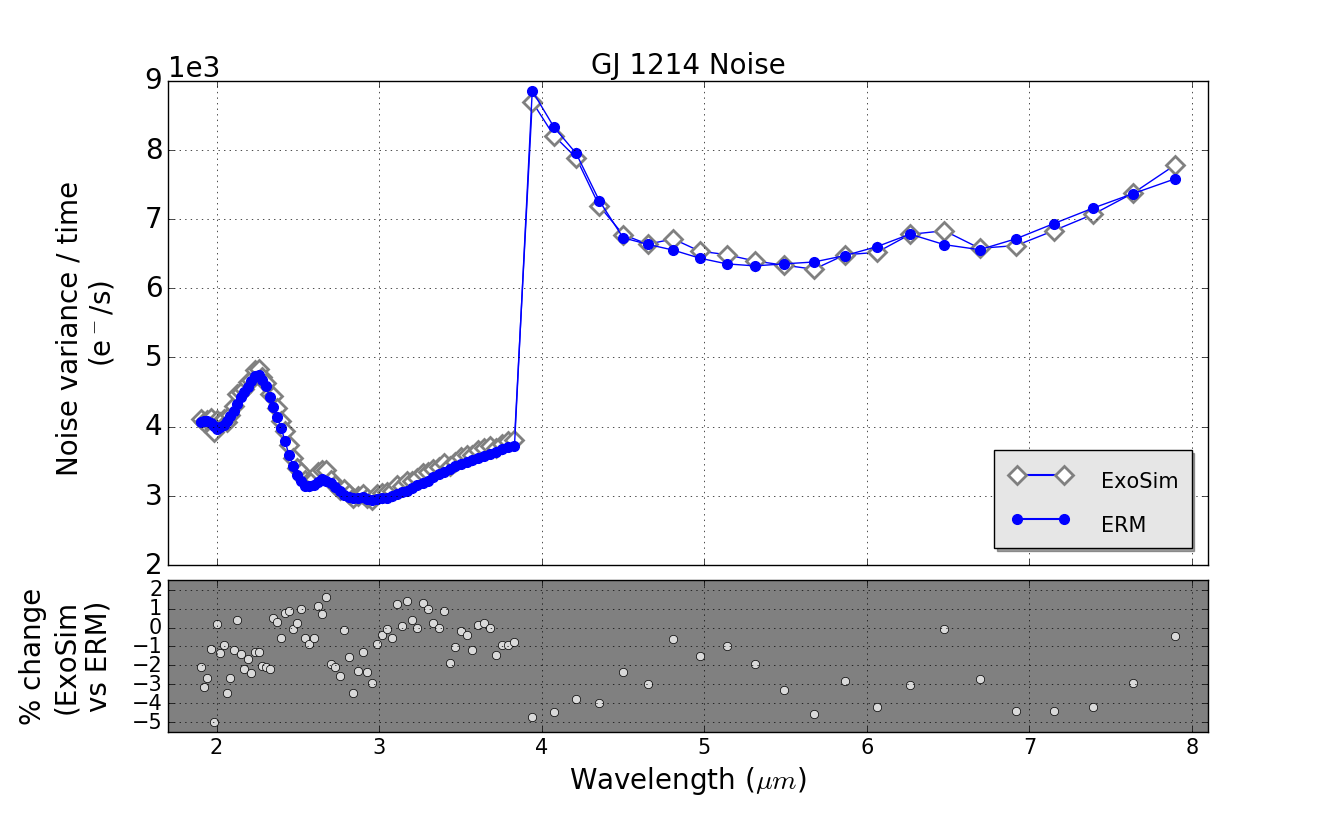}
\caption[]{ExoSim and ESA radiometric model cross-validation test. Top: signal per unit time.  Bottom: noise variance per unit time. Results are binned to R=100 for 1.9-3.9 \textmu m and R=30 for 3.9-7.8 \textmu m.  The ExoSim noise results show the average for 50 realizations.  Grey plots show percent difference of ExoSim from the prediction.}
\label{Chapter2:Figure:esa}
\end{center}
\end{figure}

The same stellar parameters and Phoenix models were used in both simulators for the stars 55 Cancri (a bright source) and GJ 1214 (a dim source).  

An early iteration of the ARIEL instrument description was modeled in each simulator.  This was a modified Offner grating design with 2 infra-red channels between 1.9 and 7.8  \textmu m, with the final spectra binned to R=100 and R=30 in each channel respectively. This design was from the ARIEL proposal document \citep{Proposal}, and does not represent the Phase A design of the instrument \citep{Tinetti2018}\footnote{Notably the dark current assumed for this early iteration was high compared to the current design.}. However, for the purpose of cross-validating the two simulation tools this approach is adequate.

Noise sources included in both simulators were the photon noise from the target star and the dark current shot noise. In ExoSim the average noise variance per unit time per spectral bin was obtained over 50 realizations of a noisy out-of-transit observation of each star. Each observation consisted of 500 exposures.  The images were processed including aperture mask of width 2.44 f$\lambda$ (to minimise dark current noise)\footnote{Processing involved correlated-double-sampling, background subtraction, aperture masking, extraction of 1-D spectra from each 2-D image, binning into spectral-resolution-element-sized bins, and obtaining timelines of counts for each bin. The masks were sized to the diameter of the Airy disc which is adequate for a diffraction limited focal plane.}.  The ERM performed an equivalent calculation of photon noise and dark current noise, assuming the same aperture. The final results are shown in Figure \ref{Chapter2:Figure:esa}, bottom.  There is good agreement, with the ExoSim average noise variance always within 5.2\% and 4\% of the ERM for 55 Cancri and GJ 1214 respectively.

To compare the signal obtained in each simulator, noiseless simulations were run in both ExoSim and the ERM.  No apertures were used or assumed.  The signal per unit time per spectral bin was obtained for both 55 Cancri and GJ 1214 using each simulator.
The results are shown in Figure \ref{Chapter2:Figure:esa}, top.  The ExoSim signal is always within 2\% of the ERM for 55 Cancri, and within 4\% for GJ 1214.

\subsection{ExoSim vs Hubble WFC3}
\label{subsubsection:vs Hubble}

We next validated ExoSim against Hubble WFC3 measurements by implementing a WFC3 IR simulation within ExoSim.  ExoSim simulated a primary spectroscopic transit of the super-Earth, GJ 1214b.  The results from ExoSim were compared to results from two published transit spectroscopy studies with WFC3 IR that observed GJ 1214b: \cite{Berta2012} (hereafter B12) observing in staring mode, and \cite{Kreidberg2014} (herafter K14) observing in spatial scanning mode.  GJ 1214 was simulated with a PHOENIX model spectrum (T=3000 K, log$g$=5.0, [Fe/H]=0) and all remaining stellar and planet parameters obtained via the Open Exoplanet Catalogue \citep{Rein}. 

The Hubble WFC3 IR instrument was modelled in ExoSim using transmissions obtained from publically available $synphot$ files \citep{synphot}. Other instrument parameters were obtained or derived from the WFC3 data and instrument handbooks \citep{WFC3data, WFC3}. These included a reciprocal linear dispersion of 0.00025 \textmu{}m/\textmu{}m, plate scale of 0.13 mas/pixel, dark current of 0.048 e$^-$/s, and a read noise of 14.7 e$^-$ per pixel. The $f$-number used to generate the PSFs is adjusted per wavelength to match the model PSF FWHM values in \cite{WFC3}. In the current model we have not taken into account the small variations in the wavelength solution with row due to the geometric distortion of the focal plane.  

\begin{figure}
\begin{center}
\includegraphics[width=1.0\textwidth]{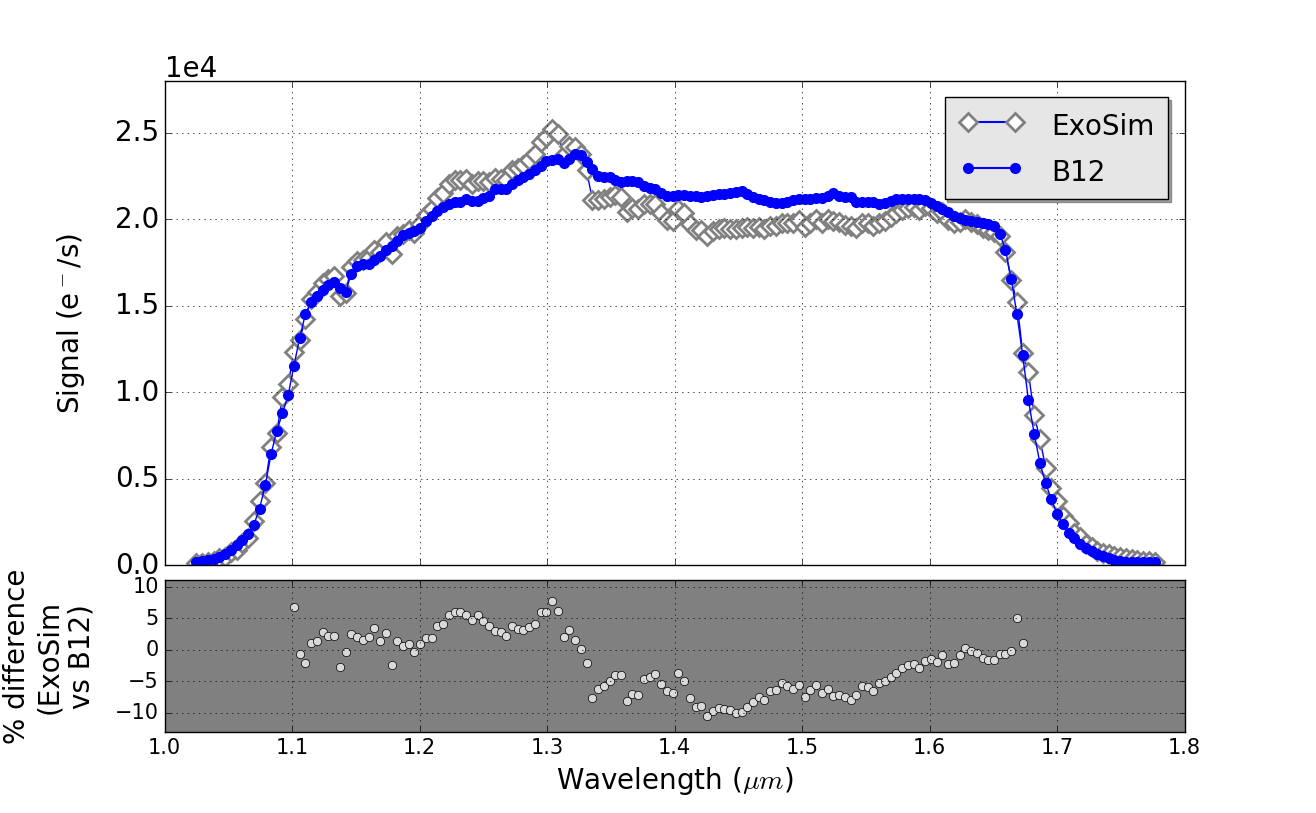}
\caption[Comparison of focal plane spectrum from ExoSim and \cite{Berta2012}.] {Comparison of focal plane spectrum from ExoSim and B12 showing photoelectron counts per pixel column per second.  Grey plot shows percent deviation of ExoSim from B12, over the wavelength range 1.10-1.67 microns.}
\label{Chapter2:Figure:berta}
\end{center}
\end{figure}

Firstly, we compare the focal plane spectral image count rates, in e$^-$/s per pixel column, from ExoSim to those published in B12 (Figure \ref{Chapter2:Figure:berta})\footnote{For comparisons with published charts, we use WebPlotDigitizer \citep{Rohatgi} to extract data points.
We assume the uncertainties arising from this extraction are small compared to the differences between the comparison data sets as evident from Figure \ref{Chapter2:Figure:berta}.}. Over the range 1.10-1.67 \textmu m we find that the ExoSim spectrum is sometimes higher and sometimes lower than the B12 spectrum, averaging 2\% lower, with a peak-to-peak variation of +8 to -11\%.  Considering that a model is being compared to the real star and instrument, the spectral shape and count rates are remarkably similar.  Integrated over all pixel columns, the total photoelectron count is 2.665x$10^6$ e$^-$/s for ExoSim, compared to 2.707x$10^6$ e$^-$/s for B12, a 1.6\% difference.

\begin{figure}
\begin{center}
\includegraphics[width=1\textwidth]{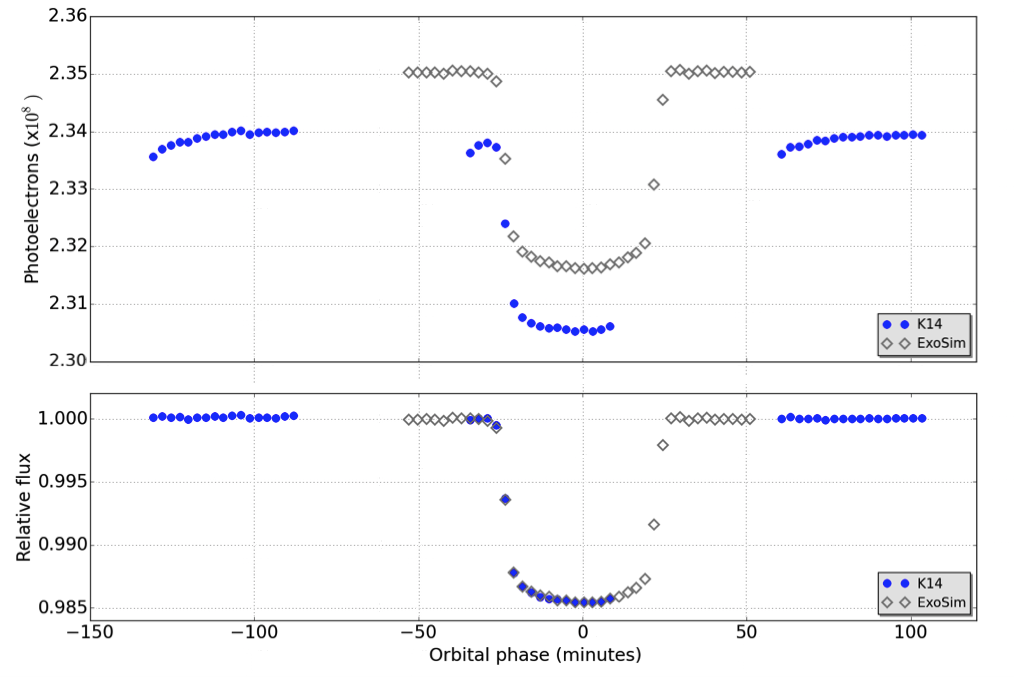}
\caption[Comparison of white light curves from ExoSim and \cite{Kreidberg2014}.]{Comparison of white light curves from ExoSim and K14.  Shown are the results from one of 20 ExoSim realizations.  K14 data obtained by resampling published graph with WebPlotDigitizer \citep{Rohatgi}. Top: absolute photoelectron counts.  Bottom: normalized light curves (with systematic correction in K14). Only orbits 2-4 are shown from K14. }
\label{Chapter2:Figure:kreidberg}
\end{center}
\end{figure}

Next, ExoSim was compared to results from K14. The ExoSim simulation was performed with observational parameters closely matched to those used in K14: spatial scanning at 12$^{\prime\prime}$/s, 90 s integration time per exposure,\footnote{Compared to 88.1 s in K12. It is unlikely this 2\% difference will significantly affect the comparison.}, 160 s cadence and 12 subexposures per exposure (where a subexposure is the difference image between adjacent NDRs). The count per exposure was obtained by summing the subexposure counts.  The characteristic `ramp' systematic due to detector persistence, and gaps in data due to Earth occultation were not simulated.  ExoSim utilized the same linear wavelength-dependent limb-darkening coefficients obtained in K14\footnote{Those obtained after using the `divide-white' systematic correction method to eliminate the `ramp' systematic.}, with the average (0.2674) used outside the published wavelength range. A flat planet transmission spectrum (consistent with known results for this planet) was used, with $(R_p/R_s)^2=0.0135$. 20 realizations of this simulation were performed.

We first compare the absolute photoelectron counts in the `white light' curves, i.e. the full array counts per exposure. 
We find that the average out-of-transit photoelectron count from ExoSim to be 2.35 x $10^8$ e$^-$, compared to 2.34 x 10$^8$ e$^-$ from orbits 2 and 4 in K14. ExoSim is thus within 0.5\% of the K14 value.  This is within the reported 1\% peak-to-peak stellar variability for the parent star in the visual range \citep{Berta2011}.  One of the 20 ExoSim realizations is shown along side the K14 data in Figure \ref{Chapter2:Figure:kreidberg}, top. 

Next we compare the white light curves from ExoSim normalised to the out-of-transit signal, with the example from K14.  In K14, in addition to out-of-transit normalisation, the `ramp' systematic is also detrended (Figure \ref{Chapter2:Figure:kreidberg}, bottom).  The photometric noise on the residuals matches closely: 70 ppm reported in K14, and $68\pm7$ ppm from ExoSim\footnote{This is the standard deviation of all residuals after a curve fitting each white light curve with a Mandel-Agol model with a fixed linear limb darkening coefficient of 0.2674.}. Comparing the K14 transit curve to one of the 20 ExoSim simulations shown in Figure \ref{Chapter2:Figure:kreidberg} (bottom), visually there is a marked similarity in the transit profiles. 

To further evaluate the similarity of the transit light curves, the 17 data points from K14 orbit 3 (the partially transiting portion of the normalised white light curve) were compared to the 17 time-matched data points in each of the 20 ExoSim simulations, using the two-sample Kolmorov-Smirnoff (K-S) test. The two-sample K-S test is used to test the null hypothesis that the two samples come from the same distribution. The test was performed for each of the 20 ExoSim simulations. An average K-S statistic was obtained of $0.16\pm0.04$, and an average p-value of $0.9\pm0.1$.  Assuming a significance level of 5\% ($\alpha = 0.05)$ for rejection of the null hypothesis, these results fail to reject the null hypothesis that the two distributions are different, and are thus supportive of the similarity between the ExoSim and K14 light curves.

We have therefore found a number of metrics in good agreement between ExoSim and the real data from these two studies. This gives us additional confidence in the accuracy of the complete end-to-end ExoSim simulation.

\section{Conclusions}

In this paper we presented ExoSim, a generic simulator of exoplanet transit spectroscopy observations.  ExoSim has been extensively validated and its end-to-end simulation is able to reproduce results obtained from an existing instrument, the Hubble WFC3 IR instrument.  ExoSim is accurate to within about 5\% of most comparisons. This gives confidence in the further use of ExoSim to simulate signal and noise in a variety of instruments and for many different purposes.  We also have demonstrated in this paper how ExoSim can be used to model different instruments, both future and existing.  ExoSim has already been applied in the performance evaluation and design of the ARIEL instrument\citep{Sarkar2017}, and in finding the noise impact from stellar pulsations and granulations \citep{Sarkar2018}. ExoSim has played a key role in validating the ARIEL mission science feasibility \citep{Tinetti2018,Pascale2018,Zingales2018,Edwards2019a}. It has also been used to assess noise and feasibility in the Twinkle space mission \citep{Edwards2019b} and the EXCITE balloon mission concept \citep{Nagler2019}.  A derivative of ExoSim, an independent development which has been modified and optimized for the James Webb Space Telescope, called JexoSim, has also been produced \citep{Sarkar2019}.

With its dedicated star spot simulator, ExoSim can be used to assess the impact of spots and faculae on transit spectroscopy observations. Since ExoSim can also model existing instruments it can potentially be used to verify the error bars obtained from previously published studies. ExoSim, through its data products, can be used to test the development of data reduction pipelines. ExoSim is thus a highly versatile tool that will aid the future development of exoplanet transit spectroscopy. 

\begin{acknowledgements}
We appreciate the help of G. Morello (CEA-Saclay) in providing limb darkening coefficients for the ExoSim database. This work has been supported by ASI grant n. 2018.22.HH.O and UKSA grant ST/S002456/1. This project has received funding from the European Research Council (ERC) under the European Union’s Horizon 2020 research and innovation programme (grant agreement No: 758892, ExoAI). 

\end{acknowledgements}

\bibliographystyle{mnras}      
\nocite{*}

\bibliography{example.bib}

\end{document}